\documentclass[11pt]{article}
\usepackage[margin=1in]{geometry}
\usepackage{algorithmic} %For computer algorithm code in LATEX
\usepackage{algorithm} %For algorithm box around the algorithmic
\usepackage{amsthm}
\usepackage{amsmath}
\usepackage{amssymb}
\usepackage{bbm}
\usepackage{graphicx}
\usepackage{xcolor}
\usepackage{caption}
\usepackage{setspace}
\usepackage{threeparttable}
\usepackage{endnotes}
\usepackage{array}
\usepackage{xspace}
\usepackage[american]{babel}
\usepackage{lscape}
\usepackage{lipsum}
\usepackage{listings}
\usepackage{ifthen}
\usepackage{csquotes}
\usepackage{subcaption}

\usepackage{multirow}

\usepackage[style=apa,sortcites=true,sorting=nyt,backend=biber, uniquename=false]{biblatex}

\newcommand{\indep}{\rotatebox[origin=c]{90}{$\models$} }

\DeclareMathOperator*{\median}{median}
\newcommand{\EXP}{E}
\newcommand{\VAR}{Var}

\newcommand{\cond}{\, | \,}
\newcommand{\con}{ ; }
\newcommand{\R}{\mathbbm{R}}
\newcommand{\ind}{\mathbbm{1}}

\newtheorem{assumption}{Assumption}

\newcommand{\PreserveBackslash}[1]{\let\temp=\\#1\let\\=\temp}
\newcolumntype{C}[1]{>{\PreserveBackslash\centering}p{#1}}
\newcolumntype{L}[1]{>{\PreserveBackslash\raggedright}p{#1}}
\newcolumntype{R}[1]{>{\PreserveBackslash\raggedleft}p{#1}}

\newcommand{\LSS}{^{(-k)}}

\usepackage{ragged2e}
\usepackage{multirow}
\usepackage{subcaption}
\usepackage{amsmath,amssymb,xcolor}
\usepackage{pgf,tikz}
\usetikzlibrary{arrows,shapes.arrows,shapes.geometric,shapes.multipart, decorations.pathmorphing,positioning,shapes.swigs,shapes,decorations,arrows,calc,arrows.meta,fit,positioning}
\tikzset{
-Latex,auto,node distance =1 cm and 1 cm,semithick ,
state/.style ={ellipse, draw, minimum width = 0.7 cm},
point/.style = {circle, draw, inner sep=0.04cm,fill,node contents={}},
bidirected/.style={Latex-Latex,dashed},
el/.style = {inner sep=2pt, align=left, sloped},
styRectDef/.style = {rectangle, rounded corners, draw=black, inner xsep=6mm, inner sep=3mm}
}

\usepackage{tikz,pgf, float}
\usepackage{centernot}

\usepackage{fancyhdr}
\theoremstyle{plain}

\newtheorem{theorem}{Theorem}[section]

\DeclareLanguageMapping{american}{american-apa}
\usepackage{accents}

\theoremstyle{plain}

\makeatletter
\newcommand{\algorithmfootnote}[2][\footnotesize]{%
\let\old@algocf@finish\@algocf@finish% Store algorithm finish macro
\def\@algocf@finish{\old@algocf@finish% Update finish macro to insert "footnote"
\leavevmode\rlap{\begin{minipage}{\linewidth}
#1#2
\end{minipage}}%
}%
}

\DeclareCiteCommand{\nparencite}
{%
\usebibmacro{cite:init}%
\usebibmacro{prenote}}
{\usebibmacro{citeindex}%
\printtext[bibhyperref]{\usebibmacro{cite}}}
{\multicitedelim}
{\usebibmacro{postnote}}

\makeatother

\usepackage{setspace}

\newcommand\numeq{\addtocounter{equation}{1}\tag{\theequation}}
\usepackage{cancel}

\begin{document}

\title{Separable Effects in Four-Arm and Two-Arm Designs}
\author{
Chan Park$^{a}$, Youmi Suk$^{b}$\\[0.25cm]
\makebox[1cm][c]{{\footnotesize $^{a}$Department of Statistics, University of Illinois Urbana-Champaign}}\\
\makebox[1cm][c]{{\footnotesize $^{b}$Department of Human Development, Teachers College Columbia University}}\\
\makebox[1cm][c]{{\footnotesize Chan Park and Youmi Suk contributed equally to this manuscript as co-first authors.}}
}
 \date{}
  \maketitle
\begin{abstract}
Robins and Richardson (2010) reformulated mediation analysis by decomposing treatments into multiple components and examining separable effects of each component. While this approach is increasingly popular, existing work has analyzed ``two-arm'' data, where components are strictly bundled and manipulated simultaneously. However, in practice, four-arm data where components are assigned independently are often available. For example, testing accommodations might strictly bundle extra time with a separate session or allow them to be assigned separately. To address this distinction, we propose a general framework for analyzing separable effects in four-arm and two-arm designs. This framework provides distinct identification and estimation strategies for each design. For estimation, we utilize efficient influence function estimators coupled with machine learning and cross-fitting techniques. Additionally, we introduce two falsification tests for key identification assumptions required in the two-arm design by leveraging four-arm data. We investigate the performance of the proposed estimators via a simulation study and demonstrate their application by studying the effect of extended time accommodations using data from the National Assessment of Educational Progress. Ultimately, this separable effects analysis enables practitioners to clearly communicate underlying mechanisms and derive informative policy recommendations.
\end{abstract}
\noindent%

\newpage

\section{Introduction}

Standard causal mediation analysis typically focuses on the natural (or pure) direct and natural indirect effects, or the controlled direct effect \parencite{robins1992identifiability, pearl2001direct}. These estimands require hypothetical interventions on mediators and rely on ``cross-world'' independence assumptions (i.e., assumptions with simultaneous interventions on treatment $A=a$ and $A=a'$). In practice, direct interventions on mediators are often infeasible, and cross-world assumptions are neither testable nor experimentally implementable \parencite{robins2022interventionist}. To address these limitations, Robins and his colleagues \parencite{robins2010alternative, robins2022interventionist} reformulated mediation analysis in terms of interventions on \textit{treatment components} rather than on mediators, where treatment (e.g., cigarettes) is decomposed into multiple treatment components (e.g., nicotine and non-nicotine components). While this decomposition has been increasingly adopted in the literature \parencite[e.g.,][]{Stensrud2022, shpitser2017modeling, park2024proximal, suk2026identifying}, all existing work has focused on data where treatment components are manipulated simultaneously under the same treatment status, so-called ``two-arm'' data, rather than ``four-arm'' data, where each component is manipulated separately. The overarching goal of this study is to propose a general framework for treatment decomposition in four-arm and two-arm designs.  

%[Treatment Decomposition Example; why it is needed.]
Treating a complex treatment as a single intervention is often problematic because it typically represents a composite bundle rather than a single variable. For example, cigarettes are a bundle of nicotine and non-nicotine components; obesity can be decomposed into diet and exercise components \parencite{pearl2018does, Hern_n_2008}, and as will be detailed in our motivating example (see Section \ref{sec:mot_ex}), extended time accommodations (ETA) can be decomposed into extra time and separate session components. However, treatment decomposition goes beyond simply viewing a treatment as a bundle because it requires each component to operate on the outcome via distinct causal pathways. Specifically, one component should directly affect the outcome, while another affects it indirectly through a mediator. In the context of smoking, it is assumed that the nicotine component of cigarettes directly affects heart attack, whereas the non-nicotine component of cigarettes affects it via hypertension. By decomposing components this way, researchers can estimate \textit{separable effects} of each component, instead of relying on natural direct and indirect effects. This approach facilitates clearer communication with subject matter experts due to the estimands' experimental interpretability \parencite{robins2022interventionist, robins2010alternative}. 

Although treatment decomposition has gained attention in statistics, biostatistics, and epidemiology \parencite[e.g.,][]{robins2010alternative, shpitser2017modeling, park2024proximal, Stensrud2022}, its application to educational or psychological data has not been explored, with the notable exception of \textcite{suk2026identifying} that investigated decomposition for non-manipulable treatment (e.g., sex, race). Furthermore, to the best of our knowledge, all prior studies have used two-arm data, largely because four-arm data are rarely available in practice. In two-arm designs, identification of separable effects requires researchers to hypothesize the presence of a four-arm trial and relies on the so-called \textit{dismissible conditions}, which are not testable in two-arm settings; see Section \ref{sec:sep_two} for more details. Fortunately, our motivating example of ETA was administered under a four-arm design, providing a unique opportunity to demonstrate the utility of treatment decomposition. However, there is an open question as to how separable effects can be identified and estimated using four-arm data, and how these results compare to the standard two-arm case. 

The main goal of this paper is to establish a general framework for separable effects in four-arm and two-arm designs. We provide distinct identification and estimation strategies within each design. For estimation, we utilize efficient influence function (EIF) estimators coupled with machine learning and cross-fitting techniques. Additionally, we leverage four-arm data to introduce two falsification tests for the key identification assumptions required in the two-arm case. We also investigate the performance of the proposed methods via a simulation study and demonstrate their application by evaluating the separable effects of ETA using data from the National Assessment of Educational Progress (NAEP).

The remainder of the paper is organized as follows. Section \ref{sec:mot_ex} illustrates our motivating example, and Section \ref{sec:review} reviews relevant literature and outlines our specific contributions. Sections \ref{sec:sep_four} and \ref{sec:sep_two} discuss the identification and estimation strategies for separable effects in four-arm and two-arm designs, respectively. Section \ref{sec:falsification} introduces two falsification tests using direct and indirect approaches. Section \ref{sec:simu} details the design and results of our simulation study, and Section \ref{sec:ETAdata} presents a real-data application using NAEP data. Section \ref{sec:con} concludes with a discussion of this study's implications and future directions.

\section{Motivating Example}\label{sec:mot_ex}

Testing accommodations are changes in test administration procedures while maintaining the integrity of test content. These accommodations are essential interventions that allow students with disabilities or English language learners to demonstrate their abilities during a test. In particular, the ETA is the most frequently provided testing accommodation, and it offers additional time for test-takers to complete a test \parencite{lovett2010extended, suk2022ordinal}. In some testing settings, however, the ETA is always administered in a separate testing session. Consequently, the ETA's effect is not simply the effect of providing extra time, and it is entangled with the effects of both extra time and the separate session. Treating ETA as a single intervention is therefore ill-defined. A better analytic approach would be to decompose ETA into its two components in order to estimate the separable effects of each component. 

\begin{figure}[ht]
\centering
\begin{tikzpicture}
\tikzset{line width=1pt, outer sep=1pt,
ell/.style={draw,fill=white, inner sep=3pt,
line width=1pt},
swig vsplit={gap=5pt,
inner line width right=0.5pt},
ellempty/.style={fill=white}};
\node[name=A,ell,shape=ellipse] at (0,0) [label = below:{\footnotesize ETA} ] {$A$}; 
\node[name=A_Y,ell,shape=ellipse] at (2.0,0) [label = below:{\footnotesize Separate session} ] {$A_Y$};
\node[name=A_M,ell,shape=ellipse] at (1.7,0.85) [label = above:{\footnotesize Extra time} ] {$A_M$};
\node[name=D,ell,shape=ellipse] at (4.5,2) [label = below:{\footnotesize {\shortstack{Item\\access}}} ] {$M$};
\node[name=Y,ell,shape=ellipse] at (7, 0) [label = below:{\footnotesize Test score} ] {$Y$};
\node[name=X,ell,  shape=ellipse] at (4,3.5) {X};

\draw [->, line width=0.75pt] (A) to (A_Y);
\draw [->, line width=0.75pt] (A) to (A_M);
\draw [->, line width=0.75pt] (D) to (Y);
\draw [->, line width=0.75pt] (A_Y) to (Y);

\draw [->, line width=0.75pt] (X) to[in=100, out=190] (A);
\draw [->, line width=0.75pt] (X) to[in=90, out=-10] (Y);
\draw [->, line width=0.75pt] (X) to[in=110, out=-80] (D);

\draw [->, line width=0.75pt] (A_M) to (D);

\end{tikzpicture}
\caption{Directed acyclic graph illustrating the decomposition of the extended time accommodation (ETA) $A$ into a separate session $A_Y$ and extra time $A_M$, along with item access $M$, test score $Y$, and pre-treatment covariates $X$.}\label{fig:DAG__ETA}
\end{figure}
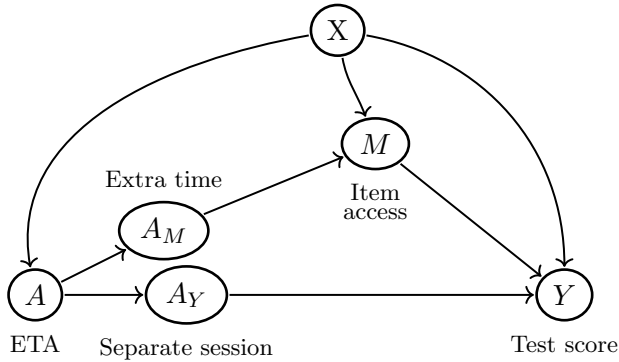

Figure \ref{fig:DAG__ETA} illustrates this treatment decomposition, where ETA ($A$) is decomposed into two binary components: extra time ($A_M \in \{0, 1\}$) and a separate session ($A_Y \in \{0, 1\}$). Testing accommodations are designed with specific cognitive purposes, so each component operates on the outcome $Y$ via distinct causal pathways. For example, the extra time component allows test-takers to actually reach and attempt more items in order to complete the assessment \parencite{lovett2010extended}. We model this as an increase in item access, which serves as a mediator, represented by the path from $A_M$ to $M$ in Figure \ref{fig:DAG__ETA}. In contrast, the separate session component is designed to minimize environmental distractions for test-takers, directly improving their focus and accuracy \parencite{weis2020separate}.\footnote{While a quiet, separate room might marginally increase the number of items a student attempts, extra time serves as the main mechanism driving item access.} We model this as a direct effect of $A_Y$ on $Y$. In practice, the administration of these components varies across testing contexts. When extra time is always bundled with a separate session, it produces two-arm data, that is, $(a_M, a_Y) \in \{(0,0), (1,1)\}$, where $A \equiv A_M \equiv A_Y$. In contrast, for other testing settings (e.g., the 2017 NAEP assessment), extra time can be provided with or without a separate session, yielding four-arm data: $(a_M,a_Y) \in \{(0,0),(0,1),(1,0),(1,1)\}$.    

Indeed, many interventions in education and psychology are not monolithic but rather comprise multiple components, each of which potentially exhibits different causal pathways to the outcome. For example, \textit{scholarship} in education policy is often treated as a single intervention, but actually consists of a monetary component and a support component (e.g., mentoring). Similarly, \textit{flipped classroom learning} in educational technology is frequently compared to traditional lecture teaching, but this intervention changes both when content is offered and how class time is used. That is, it consists of an instructional redesign component (e.g., pre-recorded lectures) and an interactive environment component (e.g., active learning activities in class). Furthermore, in clinical psychology,  \textit{cognitive behavioral therapy} is also a ``package'' intervention that consists of a behavioral activation component (e.g., physical activity) and a cognitive restructuring component (e.g., Socratic questioning). When assessing the effectiveness of such programs, it is crucial to decompose the intervention and quantify the contribution of each component, rather than using an ill-defined, composite treatment.

\section{Prior Work and Our Contributions}\label{sec:review}

\subsection{Causal Mediation Frameworks}\label{sec:review_fun}

The standard causal mediation framework \parencite{robins1992identifiability, pearl2001direct} seeks to decompose the total effect of a treatment on an outcome into direct and indirect components. We let $A \in \{0,1\}$ denote a binary treatment variable, where $A=1$ indicates receipt of treatment; let $M  \in \R^{d}$ (which may be vector-valued) denote the mediator, and let $Y \in \R$ denote the outcome. $M^{a} \in \R^{d}$ denotes the potential/counterfactual value of the mediator that would be observed if the treatment $A$ were set to $a$, and $Y^{a,m}$ denotes the potential outcome if $A$ were set to $a$ and the mediator $M$ were set to $m$. The nested counterfactual $Y^{a, M^{a'}}$ represents the potential outcome if treatment $A$ were set to $a$ but the mediator were set to the natural value it would have taken under treatment level $a'$. 

The typical estimands of interest in causal mediation analysis are the natural direct effect (NDE) and natural indirect effect (NIE)\footnote{\textcite{robins1992identifiability} used different nomenclature to refer to these effects; specifically, they termed these quantities the ``pure direct effect'' and the ``total indirect effect.''}
, which are defined as: 
\begin{align*}
\text{NDE}(a) &= E[Y^{1, M^{a}} - Y^{0, M^{a}}], \\
\text{NIE}(a) &= E[Y^{a,M^{1}} - Y^{a,M^{0}}].
\end{align*}
In words, the NDE measures the effect of the treatment (e.g., ETA) on the outcome (e.g., test scores) when the mediator (e.g., item access) is held fixed at the level it would naturally attain under a certain treatment status $A=a$. Intuitively, it represents the expected change in the outcome when switching the treatment from $A=0$ to $A=1$, while blocking the treatment’s influence on the mediator by fixing it at $M^{a}$. In contrast, the NIE captures the effect of the treatment that operates through the mediator; it considers the treatment effect on the outcome where the treatment is held fixed at a set level $A=a$, while the mediator changes from its potential value under control ($M^0$) to its potential value under treatment ($M^1$). 

Identification of the NDE and NIE requires the following cross-world independence assumption \parencite{pearl2001direct, Andrews2021}:
\begin{align}
    Y^{a,m} \indep \ M^{a'} \ , \quad \forall (a,a',m) \ .
    \label{eq-1}
\end{align}
This assumption states that the potential outcome $Y^{a,m}$ (i.e., the potential outcome under assignments $A=a$ and $M=m$) is independent of the potential mediator value $M^{a'}$, which is realized in a different hypothetical world where the treatment is set to $A=a'$. It imposes an independence relationship between counterfactual quantities defined under different treatment assignments. Unfortunately, assumption \eqref{eq-1} cannot be verified, even in well-conducted randomized experiments, because the treatment variable $A$ simultaneously takes the distinct values $a$ and $a'$. This is physically impossible. As a result, the standard mediation framework is often controversial due to the untestability of cross-world independence.

Alternative approaches to the causal mediation framework have been developed to avoid using nested counterfactuals under different treatment values (e.g., $Y^{a,M^{a'}}$) and the cross-world independence assumption. One such approach uses \textit{randomized interventional analogues} of indirect and direct effects, also referred to as \textit{interventional effects} \parencite{Geneletti2007identifying, vanderweele2014effect, Lok2016}. Rather than fixing the mediator at the value it would naturally attain under a particular treatment, interventional effects assign each individual's mediator value 
by drawing from its conditional distribution given the assigned treatment. This avoids cross-world counterfactuals, but the resulting estimands---defined in terms of a stochastic draw rather than a concrete value---remain difficult to communicate to practitioners or translate into policy decisions. Another alternative focuses on \textit{separable effects} \parencite{robins2010alternative}. We introduce relevant concepts in the next section, as it provides the key conceptual framework for this paper.

% One such approach uses \textit{randomized interventional analogues} of indirect and direct effects, also referred to as \textit{interventional effects} \parencite{Geneletti2007identifying, vanderweele2014effect, Lok2016}. The definition of interventional effects requires each individual to take a randomly sampled mediator value, rather than their naturally occurring value under the baseline treatment of interest. Specifically, an individual's mediator value is assigned stochastically from its conditional distribution given the assigned treatment, without the need to determine the value at the individual level. However, this approach is still conceptually difficult to communicate to practitioners. Another alternative focuses on \textit{separable effects} \parencite{robins2010alternative}. We introduce relevant concepts in the next section, as it provides the key conceptual framework for this paper.

%Rather than fixing the mediator at a specific counterfactual value for each individual, interventional effects characterize the impact of a treatment that shifts the distribution of the mediator. 

\subsection{Separable Effects}\label{sec:review_pro}

\textcite{robins2010alternative} introduced the concept of separable effects based on treatment decomposition, as an alternative framework for causal mediation analysis. Treatment decomposition divides a treatment into conceptually meaningful multiple components, each of which operates on the outcome through different causal pathways. In our motivating example, ETA $A$ is decomposed into separate session $A_Y$ and extra time $A_M$, and each component influences the outcome via distinct mechanisms.

\begin{figure}[!htbp]
    \centering
\subcaptionbox{DAG for treatment decomposition \label{fig:extDAG}}[.45\textwidth]{
\begin{tikzpicture}
\tikzset{line width=1pt, outer sep=1pt,
ell/.style={draw,fill=white, inner sep=3pt,
line width=1pt},
swig vsplit={gap=5pt,
inner line width right=0.5pt}};
\node[name=A,ell,shape=ellipse] at (0,0){$A$}; 
\node[name=A_M,ell,shape=ellipse] at (2,1){$A_M$};
\node[name=A_Y,ell,shape=ellipse] at (2,0){$A_Y$};
\node[name=X,ell,shape=ellipse] at (4,2){$M$};
\node[name=Y,ell,shape=ellipse] at (6, 0){$Y$};
\draw [->, line width=1.75pt] (A) to[in=-160] (A_M);
\draw [->, line width=1.75pt] (A) to (A_Y);
\draw [->, line width=0.75pt] (X) to (Y);
\draw [->, line width=0.75pt] (A_M) to (X);
\draw [->, line width=0.75pt] (A_Y) to (Y);
\end{tikzpicture}}
\hspace{0.1in}
    \subcaptionbox{SWIG for treatment decomposition \label{fig:extSWIG}}[.45\textwidth]{
\begin{tikzpicture}
\tikzset{line width=1pt, outer sep=1pt,
ell/.style={draw,fill=white, inner sep=3pt,
line width=1pt},
swig vsplit={gap=5pt,
inner line width right=0.5pt}};
\node[name=A,ell,shape=ellipse] at (0,0){$A$}; 
\node[name=A_M,shape=swig vsplit] at (2,1){  \nodepart{left}{$A_M$} \nodepart{right}{$a_M$} };
\node[name=A_Y,shape=swig vsplit] at (2,0){  \nodepart{left}{$A_Y$} \nodepart{right}{$a_Y$} };
\node[name=X,ell,shape=ellipse] at (4,2){$M^{a_M}$};
\node[name=Y,ell,shape=ellipse] at (6, 0){$Y^{a_Y, a_M}$};
\draw [->, line width=1.75pt] (A) to[in=-160] (A_M);
\draw [->, line width=1.75pt] (A) to (A_Y);
\draw [->, line width=0.75pt] (X) to (Y);
\draw [->, line width=0.75pt] (A_M) to (X);
\draw [->, line width=0.75pt] (A_Y) to (Y);
\end{tikzpicture}
    }
    \caption{Directed acyclic graph (DAG) and single-world intervention graph (SWIG) for treatment decomposition}\label{fig:extDAGSWIG}
\end{figure}
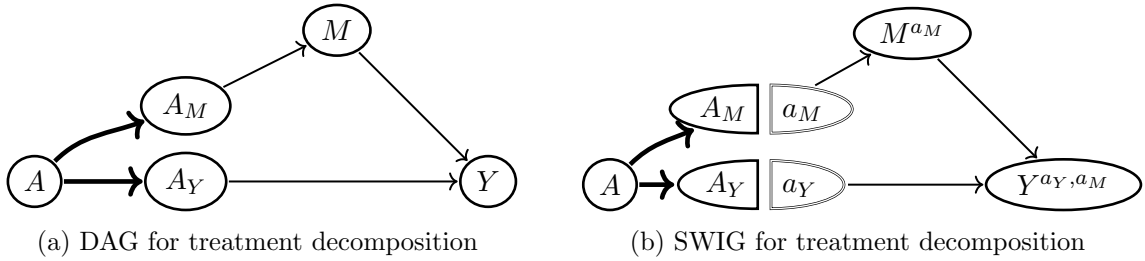

Figure \ref{fig:extDAGSWIG} visualizes this treatment decomposition using a directed acyclic graph and a single-world intervention graph (SWIG)\footnote{SWIG is a causal framework that unifies causal graphs and potential outcomes and is based on the Finest Fully Randomized Causally Interpreted Structural Tree Graph (FFRCISTG) model \parencite{robins1986ffrcistg, richardson2013swig}.}. As in many previous studies, these graphs assume a two-arm data setting, where the bold arrows from $A$ to $A_Y$ and $A_M$ represent a deterministic relationship, i.e.,  $A \equiv A_Y \equiv A_M$. Component $A_Y$ affects the outcome directly, whereas $A_M$ affects the outcome only through the mediator. Specifically,  the SWIG in Figure \ref{fig:extSWIG} splits each component node ($A_Y$ or $A_M$) into random and fixed halves, where a random half represents an observed component variable and a fixed half represents an intervention on that component. In this single world, all individuals were hypothetically assigned to the component values $A_Y=a_Y$ and $A_M=a_M$, where the potential mediator and potential outcome of interest are $M^{a_M}$ and $Y^{a_Y, a_M}$. 

In this framework, the target estimands are separable effects of each treatment component on the outcome, namely \emph{separable direct effect} (SDE) and \emph{separable indirect effect}. Formally, the SDE is defined as:
\begin{align*}
\text{SDE}(a_M) = E[Y^{1,a_M} - Y^{0,a_M}].
\end{align*}
This effect measures the effect of component $A_Y$ on outcome, holding $A_M$ constant at level $a_M$. In the ETA example, $\text{SDE}(a_M)$ represents the effect of a separate session on test scores while holding separate session status constant. Similarly, the SIE is defined as:
\begin{align*}
\text{SIE}(a_Y) = E[Y^{a_Y,1} - Y^{a_Y,0}].
\end{align*}
This effect measures the causal effect of the $A_M$ component on the outcome, holding $A_Y$ fixed at level $a_Y$. In the ETA example, $\text{SIE}(a_Y)$ represents the effect of extra time on test scores while keeping the separate session status fixed.

Unlike the estimands in traditional causal mediation analysis, the separable effects are very intuitive and require only that the interventions on $A_M$ and $A_Y$ be substantively meaningful. This allows for straightforward discussions with subject matter experts (e.g., teachers, administrators) regarding feasible experimental interventions \parencite{robins2022interventionist}. Despite these advantages, all the prior studies are based on two-arm data settings, like Figure \ref{fig:extDAGSWIG}, and research on separable effects using four-arm data is virtually non-existent.

\subsection{Our Contributions}

This paper makes four primary contributions to the causal inference and testing accommodation literature. First, we review \textit{causal identification} of separable effects under four-arm and two-arm designs. In the four-arm setting, identification follows from standard results by treating the problem as a multi-treatment setting. In contrast, identification with two-arm data requires additional assumptions \parencite[e.g.,][]{robins2010alternative, robins2022interventionist, Stensrud2021, Stensrud2022}. Specifically, prior work relies on conceptualizing a ``hypothetical'' four-arm study in which all arms are fully randomized. Here, we relax this requirement by replacing full randomization with a weaker ignorability condition (Assumption \ref{assumption-2arm-ignorability}). Under this weaker assumption, we establish that separable effects remain identifiable.

Second, we provide \textit{nonparametric estimators} for separable effects. Existing work on separable effects has focused on estimation using parametric models \parencite{robins2010alternative, robins2022interventionist, Stensrud2021, Stensrud2022, Stensrud2022CSE}, with the notable exception of \textcite{park2024proximal}. We extend this literature by developing nonparametric EIF estimators for separable effects in both four-arm and two-arm designs, grounded in semiparametric efficiency theory \parencite{BKRW1998}.

Third, we introduce \textit{two falsification tests} for the key identification assumptions in the two-arm design, namely  Assumptions \ref{assumption-2arm-isolation} and \ref{assumption-2arm-dismissible} (see Section \ref{sec:assum_two} for more details). While these assumptions are not testable using only two-arm data,  the availability of four-arm data allows us to construct falsification tests. The first test is a direct test that evaluates implied null hypotheses using regression models, while the second is an indirect test comparing separable effect estimates obtained from two-arm versus four-arm studies; see Section \ref{sec:falsification} for further details. These tests provide practical tools for assessing the credibility of the separable effects framework in applied settings.

Lastly, we derive \textit{policy implications} that are substantively meaningful and straightforward to communicate. We apply our methods to investigate the causal mechanisms of ETA on math test scores using the 2017 NAEP data. Unlike the natural direct and indirect effects used in prior work \parencite[e.g.,][]{suk2026causal}, this paper adopts a treatment decomposition perspective, where we separate ETA into extra time and a separate session, and estimate the corresponding separable direct and indirect effects. The results of our NAEP data analysis are intuitive to communicate with subject matter experts (e.g., educators, test administrators) and yield informative policy recommendations. For example, based on our empirical findings, we recommend offering extra time if students are accommodated with a separate session in order to mitigate the latter's negative effect.

%Unlike natural direct and indirect effects, separable causal effects do not rely on mediating pathways; instead, they focus on treatment components that can, in principle, be independently assigned, providing a clearer operational interpretation. To further enhance credibility, we implement the falsification tests described in Contribution 2. Taken together, this framework offers transparent, policy-relevant insights by explicitly linking specific treatment components to their outcomes, avoiding the ambiguities of mediation-based approaches and making results more interpretable for practitioners.

\section{Separable Effects in Four-Arm Designs}\label{sec:sep_four}

In this section, we discuss the identification and estimation of separable effects in four-arm designs. In these designs, all four treatment combinations $(A_Y,A_M) \in \{(0,0),(1,0),(0,1),(1,1) \}$ are fully observed, and we estimate the nuisance functions (i.e., outcome and treatment models) nonparametrically. Since this data structure allows separable effects to be framed as standard average treatment effects in a multi-categorical treatment setting, causal inference is relatively straightforward.

\subsection{Identification}\label{sec:assum_four}

Following the notation introduced in Section \ref{sec:review_pro}, we provide the conditions required to identify the separable effects.

\begin{assumption}\label{assumption-4arm}
\leavevmode
\begin{itemize}
    \item[(i)] (Causal Consistency) $Y=Y^{A_Y,A_M}$;
    \item[(ii)] (Ignorability) $Y^{a_Y,a_M} \indep (A_Y,A_M) \cond X$;
    \item[(iii)] (Positivity) For some $c>0$, we have $\Pr(A_Y=a_Y,A_M=a_M \cond X) \geq c$ for all $(a_Y,a_M) \in \{(0,0),(0,1),(1,0),(1,1)\}$.
\end{itemize}
\end{assumption}
Assumption \ref{assumption-4arm} comprises standard conditions for causal inference tailored to four-arm designs. Specifically, condition \ref{assumption-4arm}-(i) states that the observed outcome (e.g., test scores) corresponds to the potential outcome under the actually assigned treatment statuses $A_Y$ and $A_M$ (e.g., extra time and separate session statuses in the ETA setting). This implies that an individual's potential outcomes are independent of others' treatment assignments (no interference) and that there are no multiple versions of the treatment. Condition \ref{assumption-4arm}-(ii) states that, for each value of the covariates $X$ (e.g., sex, race/ethnicity, prior math ability), the treatment assignment $A_Y, A_M$ is as good as randomly assigned, and thus, independent of the potential outcome $Y^{a_Y,a_M}$. Furthermore, condition \ref{assumption-4arm}-(iii) states that the probability of receiving any of the four treatment options, i.e., the \emph{propensity score}, is strictly positive for every value of the covariates; see Chapter 3 of \textcite{HR2020} for a discussion.

It is well-known that the counterfactual mean $ \EXP [ Y^{a_Y,a_M} ]$ is nonparametrically identified under the three conditions of Assumption \ref{assumption-4arm}. For convenience, let $\nu^*(a_Y,a_M,X) = \EXP (Y \cond A_Y=a_Y,A_M=a_M,X)$ be the outcome regression model and $\pi^* (a_Y,a_M,X) = \Pr(A_Y=a_Y,A_M=a_M \cond X)$ be the propensity score model, both under four-arm designs. The counterfactual mean $ \EXP [ Y^{a_Y,a_M} ]$ is then represented in terms of $\nu^*$ and $\pi^*$ as follows:
\begin{align}
    \EXP [ Y^{a_Y,a_M} ]
    & 
    =
    \EXP 
    \bigg[ 
        \frac{\ind(A_Y=a_Y,A_M=a_M) Y }{\pi^*(a_Y,a_M,X)}
    \bigg]
     \label{eq-4arm-IPW}
    \\
    &
    =
    \EXP 
    \big[ 
    \nu^*(a_Y,a_M,X)
    \big]
    \label{eq-4arm-OR}
    \\
    &
    =
    \EXP 
    \bigg[ 
    \underbrace{
    \frac{\ind(A_Y=a_Y,A_M=a_M) \{ Y - \nu^*(a_Y,a_M,X) \} }{\pi^*(a_Y,a_M,X)}
    +
    \nu^*(a_Y,a_M,X)
    }_{\equiv \phi (O \con a_Y,a_M, \delta^*)}
    \bigg]
    \ .
     \label{eq-4arm-AIPW}
\end{align}
Equations \eqref{eq-4arm-IPW}, \eqref{eq-4arm-OR}, and \eqref{eq-4arm-AIPW} are commonly referred to as the inverse probability-weighted (IPW), outcome regression-based (OR), and augmented inverse probability-weighted (AIPW) representations, respectively. We denote the term inside the expectation of the AIPW representation as $\phi$, which represents an uncentered EIF for the counterfactual mean in the nonparametric model. We also denote the true nuisance functions as $\delta^* = (\pi^*,\nu^*)$.\footnote{Note that the AIPW representation achieves a doubly robust property. That is, if at least one of the specified nuisance functions matches its true counterpart, it correctly identifies the counterfactual mean.} 

Based on \eqref{eq-4arm-AIPW}, we identify separable effects by contrasting the representations of the counterfactual means. Specifically, our target estimands, SDE and SIE, are identified based on the AIPW/EIF-based representations:
\begin{align}
    &
    \text{SDE}(a_M) = \EXP[ 
    \phi(O \con 1,a_M, \delta^*)
    -
    \phi(O \con 0,a_M, \delta^*)
    ]
    \ , 
    \nonumber
    \\
    &
    \text{SIE}(a_Y) = \EXP[ 
    \phi(O \con a_Y,1, \delta^*)
    -
    \phi(O \con a_Y,0, \delta^*)
    ] \ .
    \label{eq-SIE-IF}
\end{align}
Here, $O=(Y,M,A_Y,A_M,X)$. Similar identification results hold for the IPW and OR representations \eqref{eq-4arm-IPW} and \eqref{eq-4arm-OR}. However, we focus on the EIF-based representations because they allow researchers to construct asymptotically normal estimators, which we will detail in the next section. 

\subsection{Estimation}\label{sec:estimation_four}

In four-arm designs, we estimate the SDE and SIE using the EIF-based AIPW estimators. In particular, this estimation strategy aligns closely with double/debiased machine learning \parencite[DML;][]{Victor2018} applied to a multi-categorical treatment.\footnote{DML provides a general framework for estimating low-dimensional causal parameters (e.g., the average treatment effect, regression coefficients) in the presence of high-dimensional or flexible nuisance functions using machine learning methods while still obtaining valid $\sqrt{n}$-consistent, asymptotically normal estimators for the target parameter \parencite{Victor2018}.} To implement this, we utilize a cross-fitting procedure \parencite{Schick1986, Victor2018}. We randomly partition the full sample into $K$-folds. For each fold $k \in \{1, \ldots, K\}$, let $\mathcal{I}^{(-k)}$ denote the estimation fold and $\mathcal{I}^{(k)}$ denote the evaluation (or held-out) fold. Using the estimation fold $\mathcal{I}^{(-k)}$, we estimate $\delta^* = (\nu^*, \pi^*)$ via flexible machine learning methods, yielding estimators $\widehat{\delta}\LSS = (\widehat{\nu}\LSS, \widehat{\pi}\LSS)$. Specifically, the propensity score $\pi^*(a_Y,a_M,X)$ can be estimated by regressing each treatment indicator $\ind(A_Y=a_Y,A_M=a_M)$ on the covariates $X$, and the outcome regression $\nu^*(A_Y,A_M,X)$ can be estimated by regressing the outcome $Y$ on the treatment components and covariates $(A_Y,A_M,X)$. In our empirical implementation, we use $K=2$ folds and fit random forests \parencite{RF} for both the outcome and treatment models, with varying numbers of trees (e.g., 500, 1,000, 1,500). We then combine the predictions from these random forests using a super learning algorithm \parencite{superlearer}.

Next, we use these estimated nuisance functions on the observations in the evaluation fold $\mathcal{I}^{(k)}$ to compute the estimated EIF $\phi$. Finally, we aggregate the resulting EIF estimates across all $K$ folds to construct the estimators of the separable effects. The resulting estimators are as follows:
\begin{align}
    & 
    \widetilde{\text{SDE}}(a_M)
    =
    \frac{1}{ N }
    \sum_{k=1}^{K}
    \sum_{i \in \mathcal{I}^{(k)} }
    \big\{ 
        \phi(O_i \con 1,a_M, \widehat{\delta}\LSS)
    -
    \phi(O_i \con 0,a_M, \widehat{\delta}\LSS)
    \big\}  \ ,
    \nonumber
    \\
    &
    \widetilde{\text{SIE}}(a_Y)
    =
    \frac{1}{ N }
    \sum_{k=1}^{K}
    \sum_{i \in \mathcal{I}^{(k)} }
    \big\{ 
        \phi(O_i \con a_Y,1, \widehat{\delta}\LSS)
    -
    \phi(O_i \con a_Y,0, \widehat{\delta}\LSS)
    \big\}  \ .
    \label{eq-4arm estimator}
\end{align} 
It can be shown that these estimators are asymptotically normal for the separable effects under standard regularity conditions about the boundedness and the $L_2$-convergence rates of $o_P(N^{-1/4})$ for the nuisance estimators. Details of these regularity conditions are provided in Supplementary Appendix \ref{app:regu_cond}.

\begin{theorem} \label{thm-4arm}
    Suppose that Assumptions \ref{assumption-4arm} and the standard regularity conditions hold. Then, we have
    \begin{align*}
        &
        \sqrt{N} \big\{ \widetilde{\text{SDE}}(a_M) - \text{SDE}(a_M) \big\}
        \stackrel{D}{\rightarrow} 
        N \big( 0, \sigma_{\text{SDE}}^2  (a_M) \big) \ , 
        \\
        &
        \sqrt{N} \big\{ \widetilde{\text{SIE}}(a_Y) - \text{SIE}(a_Y) \big\}
        \stackrel{D}{\rightarrow} 
        N \big( 0, \sigma_{\text{SIE}}^2  (a_Y) \big) \ ,
    \end{align*}
    where
    \begin{align*}
    &
    \sigma_{\text{SDE}}^2(a_M)
        =
        \VAR \big\{  \phi(O_i \con 1,a_M, \delta^* )
    -
    \phi(O_i \con 0,a_M, \delta^* )
    \big\}
    \ , 
    \\
    &
    \sigma_{\text{SIE}}^2(a_Y)
        =
        \VAR \big\{  \phi(O_i \con a_Y, 1, \delta^* )
    -
    \phi(O_i \con a_Y, 0, \delta^* )
    \big\}
    \ .
    \end{align*}
    In addition, consistent estimators of $\sigma_{\text{SDE}}^2$ and $\sigma_{\text{SIE}}^2$ are given by
    \begin{align*}
    &
    \widetilde{\sigma}_{\text{SDE}}^2(a_M)
    =
    \frac{1}{ N }
    \sum_{k=1}^{K}
    \sum_{i \in \mathcal{I}^{(k)} }
    \big\{ 
        \phi(O_i \con 1,a_M, \widehat{\delta}\LSS)
    -
    \phi(O_i \con 0,a_M, \widehat{\delta}\LSS)
    -
    \widetilde{\text{SDE}}(a_M)
    \big\}^2  \ ,
    \\
    &
    \widetilde{\sigma}_{\text{SIE}}^2(a_Y)
    =
    \frac{1}{ N }
    \sum_{k=1}^{K}
    \sum_{i \in \mathcal{I}^{(k)} }
    \big\{ 
        \phi(O_i \con a_Y,1, \widehat{\delta}\LSS)
    -
    \phi(O_i \con a_Y,0, \widehat{\delta}\LSS)
    -
    \widetilde{\text{SIE}}(a_Y)
    \big\}^2  \ .
    \end{align*}
\end{theorem}
Based on Theorem \ref{thm-4arm}, we construct confidence intervals for the separable effects. For example, a 95\% confidence interval for $\text{SDE}(a_M)$ is given by $\widetilde{\text{SDE}}(a_M) \pm 1.96 N^{-1/2} \widetilde{\sigma}_{\text{SDE}}(a_M)$.  We remark that although alternative estimators could be constructed using the IPW and OR representations, their asymptotic normality cannot generally be established when the nuisance parameter $\delta^*$ is estimated using flexible machine learning methods. In contrast, EIF-based estimators remain asymptotically normal even when $\delta^*$ is estimated nonparametrically \parencite[e.g.,][]{Victor2018}. Moreover, semiparametric efficiency theory implies that any asymptotically normal estimator of the separable effects must be asymptotically equivalent to an EIF-based estimator. Thus, this approach is the \textit{de facto} standard for obtaining asymptotically normal estimators under minimal assumptions. For these reasons, despite the mathematical simplicity of the IPW and OR representations in \eqref{eq-4arm-IPW} and \eqref{eq-4arm-OR}, we base our estimator of the separable effects on the EIF/AIPW representations.

Although Theorem \ref{thm-4arm} is established for a single $K$-fold sample split, finite-sample performance can be highly sensitive to the specific partition. To mitigate this dependence, \textcite{Victor2018} recommends a median adjustment based on multiple cross-fitting replicates. Let $\widetilde{\text{SDE}}^{[s]}$ denote the $s$-th cross-fitted estimate of the SDE, and let $\widetilde{\sigma}_{\text{SDE}}^{2,[s]}$ denote the corresponding variance estimate. The median-adjusted estimator for the SDE and its variance are defined as:
\begin{align*}
    &
    \widetilde{\text{SDE}}^{[\text{med}]}
    =
    \median_{s=1,\ldots,S} 
    \widetilde{\text{SDE}}^{[s]}
    \ , 
    &&
    \widetilde{\sigma}_{\text{SDE}}^{2,[\text{med}]}
    =
    \median_{s=1,\ldots,S} 
    \bigg[ 
        \widetilde{\sigma}_{\text{SDE}}^{2,[s]}
        +
        \big\{ \widetilde{\text{SDE}}^{[s]} - \widetilde{\text{SDE}}^{[\text{med}]} \big\}^2
    \bigg] \ .
\end{align*}
A similar procedure yields a median-adjusted estimator for the SIE. By aggregating over multiple splits, these median-adjusted estimators become less sensitive to a single sample partition. 

\section{Separable Effects in Two-Arm Designs} \label{sec:sep_two}

In this section, we discuss the identification and estimation of separable effects in two-arm designs, where the two treatment components are manipulated simultaneously under the same treatment status: $(A_Y,A_M) \in \{(0,0),(1,1)\}$. Therefore, $A_Y \equiv A_M \equiv A$. In our ETA context, this implies that students receive either both extra time and a separate session, or neither. Below, we highlight two key points. First, despite the absence of the four-arm data structure, separable effects can be identified under additional assumptions. Second, asymptotically normal estimators for these effects can be constructed even when the nuisance functions are estimated nonparametrically.

\subsection{Identification}\label{sec:assum_two}

In two-arm designs, let $O = (Y,M,A,X)$ denote the observed data, consisting of the outcome, mediator, treatment, and pre-treatment confounders. Let $Y^{a}$ and $M^{a}$ be the potential outcome and potential mediator that would be observed if the treatment were $A=a$. In addition, let $G$ denote a hypothetical four-arm study in which all four possible treatment combinations are available (e.g., extra time and separate session assignments are manipulated separately). We append ``$(G)$'' to variables to indicate that they are defined in this four-arm study. For example, $Y(G)$ and $Y^{a_Y,a_M}(G)$ denote the observed outcome and the potential outcome under $(A_Y=a_Y,A_M=a_M)$ in the hypothetical four-arm study. Likewise, $M(G)$ and $M^{a_Y,a_M}(G)$ denote the observed mediator and the potential mediator under $(A_Y=a_Y,A_M=a_M)$ in the four-arm study.

We outline the assumptions required to identify the target separable effects, SDE and SIE, in the two-arm design based on prior work \parencite[e.g.,][]{robins2010alternative, robins2022interventionist, Stensrud2021, Stensrud2022}.

\begin{assumption}\label{assumption-2arm-1}
\leavevmode
\begin{itemize}
    \item[(i)] (Causal Consistency) $Y=Y^{A}$; 
    \item[(ii)] (Ignorability) $(Y^{a},M^{a}) \indep A \cond X$;
    \item[(iii)] (Positivity) For some $c>0$, we have $f^*(A=a,M=m \cond X) \geq c$ for all $a$ and $m$, where $f^*(A,M \cond X)$ is the conditional probability density function of $(A,M)$ given $X$.
\end{itemize}
\end{assumption} 
Assumption \ref{assumption-2arm-1} can be interpreted analogously to Assumption \ref{assumption-4arm}, but in the presence of the mediating variable $M$ (e.g., item access). Additionally, in the two-arm case, we make Assumptions \ref{assumption-2arm-ignorability}, \ref{assumption-2arm-isolation}, and \ref{assumption-2arm-dismissible}, all of which are based on the variables from the hypothetical four-arm study $G$.

\begin{assumption}[Ignorability in the four-arm study $G$] \label{assumption-2arm-ignorability} \phantom{} 
\begin{align}
    %\label{eq-4arm-ign}
    (Y^{a_Y,a_M}(G), M^{a_Y,a_M}(G)) \indep (A_Y(G), A_M(G)) \cond X(G) \ .
\end{align}
\end{assumption}

Assumption \ref{assumption-2arm-ignorability} states that the variables in $G$ satisfy the ignorability condition, similar to Assumption \ref{assumption-2arm-1}-(ii). This assumption is weaker than the standard condition on $G$ used in the prior literature \parencite{Stensrud2021, Stensrud2022, Stensrud2022CSE}, which assumes that $(A_Y(G), A_M(G))$ are randomly assigned in the future four-arm study. Nevertheless, Assumption \ref{assumption-2arm-ignorability} is sufficient for establishing identification. In our ETA context, it means that in a four-arm study, where extra time and separate sessions are separately assigned, the treatment assignments are considered as if they were randomized conditional on the covariates.   
\begin{assumption}[Isolation] \label{assumption-2arm-isolation}
\leavevmode
\begin{itemize}
    \item[(i)] ($A_Y$ Partial Isolation) $M^{a_Y=1,a_M}(G) = M^{a_Y=0,a_M}(G)$. Therefore, with a slight abuse of notation, we denote $M^{a_M}(G) = M^{a_Y,a_M}(G)$;
    \item[(ii)] ($A_M$ Partial Isolation) $Y^{a_Y,a_M=1}(G) = Y^{a_Y,a_M=0}(G)$ if $M^{a_M=1}(G)=M^{a_M=0}(G)$.
\end{itemize}
\end{assumption}  

Assumption \ref{assumption-2arm-isolation}-(i) states that the $A_Y$ component does not exert its effect on the mediator $M$ in the future four-arm study. In our ETA context, it means that the separate session does not influence the number of items accessed (i.e., item access). Assumption \ref{assumption-2arm-isolation}-(ii) states that the $A_M$ component does not directly affect the outcome $Y$, but rather affects it only indirectly through the mediator $M$ in the future four-arm study. For our ETA example, it means that extra time influences the test scores only through item access. These two conditions ensure that $A_Y$ and $A_M$ have partially isolated effects on $Y$ and $M$, respectively.

\begin{assumption}[Dismissible Conditions] \label{assumption-2arm-dismissible}
\leavevmode
\begin{itemize}
    \item[(i)] $M(G) \indep A_Y(G) \cond ( A_M(G), X(G) )$;
    \item[(ii)] $Y(G) \indep A_M(G) \cond ( A_Y(G), M(G), X(G) )$.
\end{itemize}
\end{assumption}  
The first condition of Assumption \ref{assumption-2arm-dismissible} states that, in the future four-arm study, the mediator $M$ (e.g., item access) would be independent of the $A_Y$ component (e.g., separate session) conditional on $A_M$ and $X$ (e.g., the set of extra time and pre-treatment covariates). Likewise, the second condition implies that the outcome would be independent of the $A_M$-component conditional on $A_Y$, $M$, and $X$. It is important to note that neither Assumption \ref{assumption-2arm-isolation} nor Assumption \ref{assumption-2arm-dismissible} implies the other\footnote{Specifically, even if Assumption \ref{assumption-2arm-isolation} holds, Assumption \ref{assumption-2arm-dismissible} may fail, for instance, if there exists an unmeasured variable $U$ which is a common cause of $M$ and $Y$. Then, even if $(A_Y,A_M)$ is randomized, $M$ and $Y$ are confounded due to the unmeasured confounder $U$, thereby violating Assumption \ref{assumption-2arm-dismissible}. In contrast, Assumption \ref{assumption-2arm-dismissible} does not imply Assumption \ref{assumption-2arm-isolation} either. For example, suppose $A_Y$ has heterogeneous effects on $M$ (violation of $A_Y$ partial isolation), but these effects are balanced out in the population; it increases $M$ for half the population and decreases $M$ by the same amount for the other half. In this case, $A_Y$ clearly affects $M$ for individuals, thus violating Assumption \ref{assumption-2arm-isolation}-(i). However, at the population level, $M$ and $A_Y$ are statistically independent, satisfying Assumption \ref{assumption-2arm-dismissible}-(i).}, but under the FFRCISTG model \parencite{robins1986ffrcistg, richardson2013swig}, Assumption \ref{assumption-2arm-dismissible} implies Assumption \ref{assumption-2arm-isolation}.\footnote{More specifically, this relation holds only when a missing arrow represents an individual-level exclusion restriction (i.e., no individual-level causal effect) under the FFRCISTG model.} That is, Assumption \ref{assumption-2arm-dismissible} would fail if $A_M$ exerts effects on $Y$, not mediated via $M$, or if $A_Y$ exerts effects on $M$. 
 
Under Assumptions \ref{assumption-2arm-1}--\ref{assumption-2arm-dismissible}, the counterfactual mean $ \EXP [ Y^{a_Y,a_M} ]$ is nonparametrically identified for all four treatment combinations, even though the two-arm design includes only $(A_Y=1, A_M=1)$ and $(A_Y=0, A_M=0)$. Specifically, following prior work \parencite{robins2010alternative, robins2022interventionist, Stensrud2021, Stensrud2022}, it is identifiable as:
\begin{align}  \label{eq-2arm-mediation functional}
    \EXP [ Y^{a_Y,a_M} ]   
    =
    \EXP \big[ 
    \EXP
    \big\{ \EXP(Y | A=a_Y,M,X) \cond A=a_M,X \big\}
    \big]
    \ .
\end{align}
When $a = a_Y = a_M$ in the two-arm case, the identifying formula \eqref{eq-2arm-mediation functional} reduces to $\EXP[Y^{a,a}] = \EXP[\EXP(Y \cond A=a, X)]$, which corresponds to the outcome regression-based representation in \eqref{eq-4arm-OR}. In this scenario, only Assumption \ref{assumption-2arm-1} is necessary for identification. However, to account for units with $a_Y \neq a_M$ in the future four-arm study, Assumptions \ref{assumption-2arm-ignorability}--\ref{assumption-2arm-dismissible} are required. Notably,  \eqref{eq-2arm-mediation functional} is algebraically equivalent to the identifying formula of the ``mediation functional''  $\EXP[Y^{a_Y, M^{a_M}}]$ in the framework of \textcite{pearl2001direct}. While the identifying formulas are identical between \citeauthor{pearl2001direct}'s mediation and \citeauthor{robins2010alternative}'s separable effects, the underlying causal estimands and their respective identification strategies are rooted in fundamentally different frameworks; see Section \ref{sec:review}.

As discussed in Section \ref{sec:assum_four}, we derive the EIF for the representation in \eqref{eq-2arm-mediation functional} to construct asymptotically normal estimators for the separable effects. We build on the EIFs for causal mediation analysis obtained by \textcite{TTS2012} to construct our target EIF estimators. In particular, the uncentered EIF for $\EXP[ Y^{a_Y,a_M} ]$, denoted by $\psi$, is given by:
\begin{align}
    &
    \psi(O \con a_Y, a_M, \eta^*)
    \label{eq-psi}
    \\
    &
    =
    \frac{ \ind (A=a_Y) }{ \omega^*(a_M , X) }
    \frac{ \rho^*(a_M,M,X) }{ \rho^*(a_Y,M,X) }
    \big\{ Y - \mu^*(a_Y,M,X) \big\}
    \nonumber
    \\
    &
    \qquad 
    +
    \frac{ \ind (A=a_M) }{\omega^*(a_M,X)}
    \{ \mu^*(a_Y,M,X)
    - 
    \lambda^*(a_Y,a_M,X) \} 
    +
    \lambda^*(a_Y,a_M,X) \ ,
    \nonumber
\end{align}
where $\eta^* = (\rho^*,\omega^*,\mu^*,\lambda^*)$ is the collection of the nuisance functions defined by
\begin{align*}
    & 
    \rho^*(a,M,X) = \Pr(A=a \cond M,X)
    \ , 
    &&
    \omega^*(a,X) = \Pr(A=a \cond X)
    \ , 
    \\
    &    
    \mu^*(a,M,X)
    =
    \EXP(Y \cond A=a,M,X)
    \ , 
    &&
    \lambda^*(a,a',X)
    =
    \EXP\{ \mu^*(a,M,X) \cond A=a',X \}
    \ .
\end{align*}
Here, $\rho^*(a,M,X)$ represents the conditional probability of receiving treatment $A=a$ (e.g., ETA status) given the mediator and covariates, while $\omega^*(a,X)$ represents the standard propensity score, i.e., the conditional probability of receiving treatment $A=a$ given covariates. The function $\mu^*(a,M,X)$ denotes the conditional expectation of the outcome given the treatment, mediator, and covariates, and $\lambda^*(a,a',X)$ represents the outcome regression function $\mu^*(a,M,X)$ marginalized over the conditional distribution of the mediator under treatment status $A=a'$. If $a=a_Y=a_M$, $\psi$ simplifies to:
\begin{align*}
    &
    \psi(O \con a, a, \eta^*)
    =
    \frac{ \ind (A=a) }{ \omega^*(a , X) }
    \big\{ Y - \lambda^*(a,a,X) \big\}
    +
    \lambda^*(a,a,X)  \ .
\end{align*}
Here, $\lambda^*(a,a,X) = \EXP (Y \cond A=a, X)$ reduces to the standard outcome regression. Thus, $\psi$ coincides with the AIPW representation \eqref{eq-4arm-AIPW} of the counterfactual mean under the four-arm design, but restricted to only two levels. 

% We note that $\psi$ in \eqref{eq-psi} has a multiple robustness property in the following sense. Specifically, let $\eta^\dagger = (\rho^\dagger,\omega^\dagger,\mu^\dagger,\lambda^\dagger)$ be a posited nuisance functions.  Then, we obtain $\EXP [\psi (O \con a_Y,a_M, \eta^\dagger)] = \EXP [ Y^{a_Y,a_M} ]$ if (i) either $\rho^\dagger = \rho^*$ or $\mu^\dagger = \mu^*$ and (ii) either $\omega^\dagger = \omega^*$ or $\lambda^\dagger = \lambda^*$. In other words, as long as at least one function in $(\rho,\mu)$ and one function in $(\omega, \lambda)$ match their true counterparts, the AIPW representation correctly identifies the counterfactual mean, without requiring all four functions to be correctly specified.

The separable effects in the two-arm designs are identifiable using the EIF-based representations as follows:
\begin{align}
    &
    \text{SDE}(a_M) = \EXP[ 
    \psi(O \con 1,a_M, \eta^*)
    -
    \psi(O \con 0,a_M, \eta^*)
    ]
    \ , 
    \nonumber
    \\
    &
    \text{SIE}(a_Y) = \EXP[ 
    \psi(O \con a_Y,1, \eta^*)
    -
    \psi(O \con a_Y,0, \eta^*)
    ] \ .
    \label{eq-SIE-IF-two}
\end{align}

\subsection{Estimation}\label{sec:estimation_two}

As in Section \ref{sec:estimation_four}, we construct estimators for the SDE and SIE based on the EIF-based representations in \eqref{eq-SIE-IF-two} using cross-fitting, but we target the nuisance parameter $\eta^*=(\rho^*,\omega^*,\mu^*,\lambda^*)$ instead of $\delta^* = (\nu^*,\pi^*)$.\footnote{We note that an estimator based on the three parameters $\omega^*$, $\mu^*$, and $f(M \cond A=a,X)$ can be constructed, following \textcite{TTS2012}. However, this approach requires estimating the conditional density of $M$ given $(A,X)$. Although nonparametric estimation is possible in principle, it is practically challenging, particularly when $M$ is multivariate as in our empirical data.} Specifically, we estimate the functions $\rho^*(1,M,X)$ and $\omega^*(1,X)$ by regressing $A$ on $(M,X)$ and on $X$, respectively. We estimate the outcome regression $\mu^*(A,M,X)$ by regressing $Y$ on $(A,M,X)$, and estimate the function $\lambda^*(a,A,X) = \EXP\{ \mu^*(a,M,X) \cond A,X \}$ by regressing the fitted values of $\mu^*$ on $(A,X)$. See Supplementary Appendix \ref{app:second estimator} for an example of how these functions can be estimated. The resulting EIF estimators are as follows:
\begin{align}
    & 
    \widehat{\text{SDE}}(a_M)
    =
    \frac{1}{N}
    \sum_{k=1}^{K}
    \sum_{i \in \mathcal{I}^{(k)} }
    \big\{ 
        \psi(O_i \con 1,a_M, \widehat{\eta}\LSS)
    -
    \psi(O_i \con 0,a_M,  \widehat{\eta}\LSS)
    \big\}  \ ,
    \nonumber
    \\
    &
    \widehat{\text{SIE}}(a_Y)
    =
    \frac{1}{ N }
    \sum_{k=1}^{K}
    \sum_{i \in \mathcal{I}^{(k)} }
    \big\{ 
        \psi(O_i \con a_Y,1,  \widehat{\eta}\LSS)
    -
    \psi(O_i \con a_Y,0,  \widehat{\eta}\LSS)
    \big\}  \ .
    \label{eq-2arm estimator}
\end{align}

These estimators are asymptotically normal for the target separable effects when $\eta^*$ and $\widehat{\eta}\LSS$ satisfy standard regularity conditions regarding the boundedness and the $o_P(N^{-1/4})$ convergence rates of the nuisance estimators, given in Supplementary Appendix \ref{app:regu_cond}.

\begin{theorem} \label{thm-2arm}
    Suppose that Assumptions \ref{assumption-2arm-1}--\ref{assumption-2arm-dismissible} and the standard regularity conditions hold. Then, we have
    \begin{align*}
        &
        \sqrt{N} \big\{ \widehat{\text{SDE}}(a_M) - \text{SDE}(a_M) \big\}
        \stackrel{D}{\rightarrow} 
        N \big( 0, \sigma_{\text{SDE}}^2  (a_M) \big) \ , 
        \\
        &
        \sqrt{N} \big\{ \widehat{\text{SIE}}(a_Y) - \text{SIE}(a_Y) \big\}
        \stackrel{D}{\rightarrow} 
        N \big( 0, \sigma_{\text{SIE}}^2  (a_Y) \big) \ ,
    \end{align*}
    where
    \begin{align*}
    &
    \sigma_{\text{SDE}}^2(a_M)
        =
        \VAR \big\{  \psi(O_i \con 1,a_M, \eta^*)
    -
    \psi(O_i \con 0,a_M, \eta^*)
    \big\}
    \ , 
    \\
    &
    \sigma_{\text{SIE}}^2(a_Y)
        =
        \VAR \big\{  \psi(O_i \con a_Y, 1, \eta^*)
    -
    \psi(O_i \con a_Y, 0, \eta^*)
    \big\}
    \ .
    \end{align*}
    In addition, consistent estimators of $\sigma_{\text{SDE}}^2$ and $\sigma_{\text{SIE}}^2$ are given by
    \begin{align*}
    &
    \widehat{\sigma}_{\text{SDE}}^2(a_M)
    =
    \frac{1}{ N }
    \sum_{k=1}^{K}
    \sum_{i \in \mathcal{I}^{(k)} }
    \big\{ 
        \psi(O_i \con 1,a_M, \widehat{\eta}\LSS)
    -
    \psi(O_i \con 0,a_M, \widehat{\eta}\LSS)
    -
     \widehat{\text{SDE}}(a_M)
    \big\}^2  \ ,
    \\
    &
    \widehat{\sigma}_{\text{SIE}}^2(a_Y)
    =
    \frac{1}{ N }
    \sum_{k=1}^{K}
    \sum_{i \in \mathcal{I}^{(k)} }
    \big\{ 
        \psi(O_i \con a_Y,1, \widehat{\eta}\LSS)
    -
    \psi(O_i \con a_Y,0, \widehat{\eta}\LSS)
    -
     \widehat{\text{SIE}}(a_Y)
    \big\}^2  \ .
    \end{align*}
\end{theorem}
We apply Theorem \ref{thm-2arm} to construct confidence intervals for the separable effects. Furthermore, to reduce sensitivity to any single sample partition, we employ median-adjusted estimators for both the SDE and SIE using a strategy similar to that detailed in Section \ref{sec:estimation_four}.

\section{Assessment of the Two-arm Design} \label{sec:falsification}

Assumptions \ref{assumption-2arm-isolation} (Isolation) and \ref{assumption-2arm-dismissible} (Dismissible Conditions) are essential for identifying and estimating separable effects in two-arm designs. While these assumptions cannot be tested empirically in a standard two-arm study, they become partially falsifiable if data from a future four-arm study $G$ are available. We outline two falsification tests for this assessment: one that directly evaluates the assumptions using regression models, and the other that indirectly assesses them by comparing separable effect estimates from two-arm and four-arm designs.

\subsection{A Direct Approach: Falsification Using Regression Models}\label{sec:Fal_reg}

We first consider a regression model-based approach that directly assesses Assumption \ref{assumption-2arm-dismissible}, as well as Assumption \ref{assumption-2arm-isolation} under additional conditions. Under Assumption \ref{assumption-2arm-dismissible}, the following null hypotheses are implied:
\begin{align*}    
    H_{0(i)}: \ &
    \EXP[M(G) \cond A_Y(G)=1, A_M(G)=a_M,X(G)=x ]
    \\
    & =
    \EXP[M(G) \cond A_Y(G)=0, A_M(G)=a_M,X(G)=x ]\ ,
    \\ 
    H_{0(ii)}: \ & \EXP[Y(G) \cond A_Y(G)=a_Y, A_M(G)=1,M(G)=m,X(G)=x ]
    \\
    & =
    \EXP[Y(G) \cond A_Y(G)=a_Y, A_M(G)=0,M(G)=m,X(G)=x ]  \ .
\end{align*}
We can evaluate these hypotheses using standard parametric or nonparametric regression models. While nonparametric approaches are theoretically feasible, they typically have limited power and require computationally intensive procedures; hence, we focus on a practical implementation using parametric models. For example, to test $H_{0(i)}$, one may regress $M$ on $(A_Y, A_M, X)$ using the four-arm data from $G$ as: 
\begin{align*}
    \EXP [ M(G) \cond A_Y(G), A_M(G), X(G) ] = 
    \beta_0 + \beta_{1} A_Y(G)
     + \beta_{2} A_M(G)
     + \beta_{3} X(G) \ ,
\end{align*}
and assess whether the coefficient for $A_Y$ significantly differs from zero by testing the null $\beta_{1}=0$. For our ETA example, we regress the item access variable on the extra time, separate session, and covariates in an additive manner using four-arm data. Of course, one can add more higher-order terms to this working model if needed. A similar approach applies to testing $H_{0(ii)}$, by regressing $Y$ on the appropriate set of predictors $(A_Y, A_M, M, X)$. In this way, we can find evidence of a violation of Assumption \ref{assumption-2arm-dismissible}-(i) or (ii) if $H_{0(i)}$ or $H_{0(ii)}$ is rejected. See Section \ref{sec:ETAdata} for a demonstration.

\begin{figure}[!htbp]
    \centering
\subcaptionbox{Example 1 \label{fig-Falsification A}}[.45\textwidth]{
\scalebox{1}{
\begin{tikzpicture}
\tikzset{line width=1pt, outer sep=1pt,
ell/.style={draw,fill=white, inner sep=3pt,
line width=1pt},
elldash/.style={draw,dotted,fill=white, inner sep=3pt,
line width=1pt},
swig vsplit={gap=5pt,
inner line width right=0.5pt}};
\node[name=A_M,shape=swig vsplit] at (0.5,1){  \nodepart{left}{$A_M$} \nodepart{right}{$a_M$} };
\node[name=A_Y,shape=swig vsplit] at (0.5,0){  \nodepart{left}{$A_Y$} \nodepart{right}{$a_Y$} };
\node[name=U1,elldash,shape=ellipse] at (0,3){$U_1$};
\node[name=M,ell,shape=ellipse] at (3,2){$M^{a_M}$};
\node[name=Y,ell,shape=ellipse] at (4,0){$Y^{a_Y, a_M}$};
\draw [->, line width=0.75pt] (M) to (Y);
\draw [->, line width=0.75pt] (A_M) to (M);
\draw [->, line width=0.75pt] (U1) to (M);
\draw [->, line width=0.75pt, bend right=20] (U1) to (A_M);
\draw [->, line width=0.75pt, bend right=70] (U1) to (A_Y);
\draw [->, line width=0.75pt] (A_Y) to (Y);
\end{tikzpicture}
}}%
\hspace{0.1in}
    \subcaptionbox{Example 2 \label{fig-Falsification B}}[.45\textwidth]{
\scalebox{1}{
\begin{tikzpicture}
\tikzset{line width=1pt, outer sep=1pt,
ell/.style={draw,fill=white, inner sep=3pt,
line width=1pt},
elldash/.style={draw,dotted,fill=white, inner sep=3pt,
line width=1pt},
swig vsplit={gap=5pt,
inner line width right=0.5pt}};
\node[name=A_M,shape=swig vsplit] at (0.5,1){  \nodepart{left}{$A_M$} \nodepart{right}{$a_M$} };
\node[name=A_Y,shape=swig vsplit] at (0.5,0){  \nodepart{left}{$A_Y$} \nodepart{right}{$a_Y$} };
\node[name=U2,elldash,shape=ellipse] at (4.5,3){$U_2$};
\node[name=M,ell,shape=ellipse] at (3,2){$M^{a_M}$};
\node[name=Y,ell,shape=ellipse] at (4,0){$Y^{a_Y, a_M}$};
\draw [->, line width=0.75pt] (M) to (Y);
\draw [->, line width=0.75pt] (A_M) to (M);
\draw [->, line width=0.75pt] (U2) to (M);
\draw [->, line width=0.75pt, bend left=20] (U2) to (Y);
\draw [->, line width=0.75pt] (A_Y) to (Y);
\end{tikzpicture}
}
    }
    \caption{Two examples where $H_{0(i)}$ and $H_{0(ii)}$ are not equivalent to \eqref{eq-AY partial expectation} and \eqref{eq-AD partial expectation}, respectively. The dotted nodes represent unmeasured variables.}
    \label{fig-Falsification}
\end{figure}
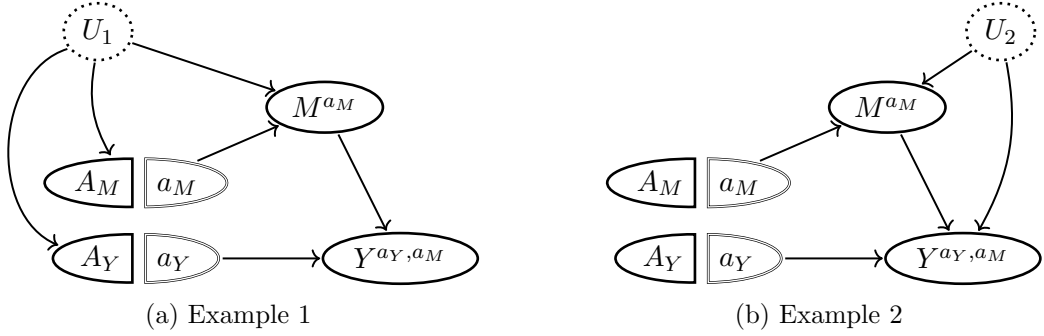

We next consider whether the regression tests above can serve as falsification tests for Assumption \ref{assumption-2arm-isolation}-(i) or (ii) ($A_Y$ partial isolation or $A_M$ partial isolation). Unlike \ref{assumption-2arm-dismissible}-(i), rejecting $H_{0(i)}$ does not provide evidence against $A_Y$-partial isolation, because other mechanisms can also produce a non-zero $\beta_1$. For instance, if there exists an unmeasured confounder $U_1(G)$ that affects both $M(G)$ and $(A_M(G), A_Y(G))$ (violating Assumption~\ref{assumption-2arm-ignorability}), then $M(G)$ and $A_Y(G)$ will be associated even when $A_Y$-partial isolation holds; see Figure~\ref{fig-Falsification A}. Consequently, rejecting $H_{0(i)}$ does not reflect a causal effect of $A_Y$ on $M$.
% For instance, suppose there exists an unmeasured confounder $U_1(G)$ that affects both $M(G)$ and $(A_M(G), A_Y(G))$ (i.e., \ref{assumption-2arm-ignorability} is violated) while $A_Y$-partial isolation still holds; see Figure \ref{fig-Falsification A} for a visual illustration. In this case, $H_{0(i)}$ is generally false because $M(G)$ and $(A_M(G), A_Y(G))$ are confounded through $U_1(G)$. Consequently, rejecting $H_{0(i)}$ does not, by itself, provide evidence against $A_Y$-partial isolation.

However, if \ref{assumption-2arm-ignorability} holds, the null $H_{0(i)}$ is equivalent to:
\begin{align}
\EXP[M^{a_Y=1,a_M}(G) \mid X(G)=x ] = \EXP[M^{a_Y=0,a_M}(G) \mid X(G)=x ]  \ .
\label{eq-AY partial expectation}
\end{align}
In this case, rejecting $H_{0(i)}$ does provide valid evidence against 
$A_Y$-partial isolation. Thus, the validity of this test depends on the plausibility of Assumption~\ref{assumption-2arm-ignorability}, which is guaranteed by design in a fully randomized four-arm study.

% Therefore, to reliably assess $A_Y$-partial isolation using this test, we must justify \ref{assumption-2arm-ignorability}; that is, we must be confident that the assignment of $(A_Y, A_M)$ is as-if random. %Ideally, this is achieved through a randomized four-arm study $G$, where the random assignment of $(A_Y, A_M)$ ensures that Assumption \ref{assumption-2arm-ignorability} holds by design.

Finally, we consider whether testing $H_{0(ii)}$ can falsify $A_M$ partial isolation (\ref{assumption-2arm-isolation}-(ii)). Similar to before, rejecting $H_{0(ii)}$ does not indicate a violation of $A_M$-partial isolation. For instance, suppose $(A_M(G), A_Y(G))$ are randomized (i.e.,  \ref{assumption-2arm-ignorability} holds by design), but there exists an unmeasured confounder $U_2(G)$ that affects both $M(G)$ and $Y(G)$ while $A_M$-partial isolation still holds; see Figure \ref{fig-Falsification B} for an illustration. In this scenario, $H_{0(ii)}$ is generally false due to the backdoor path opened by confounding through $U_2(G)$, and thus, rejecting $H_{0(ii)}$ may reflect this confounding rather than a failure of $A_M$ partial isolation. 
Unlike the case of $A_Y$ partial isolation, even if \ref{assumption-2arm-ignorability} holds, $H_{0(ii)}$ is generally not equivalent to the following condition:
\begin{align}
& \EXP[Y^{a_Y,a_M=1}(G) \mid M^{a_M=1}(G)=m, X(G)=x ]
\nonumber \\
& = \EXP[Y^{a_Y,a_M=0}(G) \mid M^{a_M=0}(G)=m, X(G)=x ] \ .
\label{eq-AD partial expectation}
\end{align} 
As a result, rejecting $H_{0(ii)}$ does not generally correspond to rejecting \eqref{eq-AD partial expectation} and thus, does not provide a valid falsification test for $A_M$-partial isolation. To achieve equivalence between $H_{0(ii)}$ and \eqref{eq-AD partial expectation}, one would need to assume both \ref{assumption-2arm-ignorability} and no unmeasured confounding between $M(G)$ and $Y(G)$. The latter is often unrealistic in practice, as it excludes any post-treatment confounders. Thus, while falsification of $A_M$ partial isolation is theoretically possible, it relies on overly restrictive assumptions and may not be well-justified in practice.

\subsection{An Indirect Approach: Falsification Using Separable Effect Estimators} \label{sec: falsification SEE}

Next, we evaluate the validity of the two-arm design by comparing its estimates with those derived from a four-arm design. In essence, a statistically significant difference between the two sets of estimates suggests that the identifying assumptions are violated. This provides an indirect assessment of those assumptions. 

We use subscripts $_{[4]}$ and $_{[2]}$ to distinguish variables from the four-arm and two-arm studies, respectively, e.g., $O_{[4]} = (Y_{[4]},M_{[4]},A_{Y,[4]}, A_{M,[4]},X_{[4]})$ and  $Y_{[4]}^{(a_Y,a_M)}$. Let $\mathcal{P}_{[2]}$ and $\mathcal{P}_{[4]}$ denote the respective study populations, with covariate densities $f_{[2]}(x)$ and $f_{[4]}(x)$, which may differ to account for potentially distinct populations across the studies. However, we assume transportability that the conditional distribution $(Y^{a_Y,a_M}, M^{a_Y,a_M}) \mid X$ is invariant across populations. This implies the conditional separable treatment effects are identical, i.e.,  $\text{SDE}_{[4]}(a_M,x)=\text{SDE}_{[2]}(a_M,x)$ and
$\text{SIE}_{[4]}(a_Y,x)=\text{SIE}_{[2]}(a_Y,x)$ where
\begin{align*}
\text{SDE}_{[\star]}(a_M, x)
&
=
\EXP [Y_{[\star]}^{1,a_M} - Y_{[\star]}^{0,a_M} \cond X_{[\star]}=x]  \ ,
\\
\text{SIE}_{[\star]}(a_Y, x)
&
=
\EXP [Y_{[\star]}^{a_Y,1} - Y_{[\star]}^{a_Y,0} \cond X_{[\star]}=x] \ , \quad \star \in \{2,4\} \ .
\end{align*}
Consequently, the separable effects must agree when the conditional separable effects are marginalized over the same covariate distribution. For instance, when marginalized over the two-arm population, we have:
\begin{align*}
    \int 
    \text{SDE}_{[4]}(a_M, x)
    \, f_{[2]}(x) \, dx
    & =
    \text{SDE}_{[2]}(a_M) \ ,
    \quad 
    \int 
    \text{SIE}_{[4]}(a_Y, x)
    \, f_{[2]}(x) \, dx
    =
    \text{SIE}_{[2]}(a_Y) \ ,
\end{align*}
where $\text{SDE}_{[2]}(a_M)$ and $\text{SIE}_{[2]}(a_Y)$ are separable effects defined over $\mathcal{P}_{[2]}$. 
Using the density ratio to adjust for population differences, this can be expressed as:
\begin{align*}
    &
    \EXP \left[ \big\{ Y_{[4]}^{1,a_M} - Y_{[4]}^{0,a_M} \big\} \frac{ f_{[2]}(X) }{ f_{[4]}(X) } \right] = \text{SDE}_{[2]}(a_M) \ , 
    \\
    &
    \EXP \left[ \big\{ Y_{[4]}^{a_Y,1} - Y_{[4]}^{a_Y,0} \big\} \frac{ f_{[2]}(X) }{ f_{[4]}(X) } \right] = \text{SIE}_{[2]}(a_Y)  \ .
    \numeq \label{eq-falsification mean}
\end{align*}
While this representation is valid for arbitrary densities and can be used for hypothesis testing below, the density ratio $f_{[2]}/f_{[4]}$ can be difficult to estimate reliably in practice. 

To tackle this problem, in this study, we utilize a simpler form in a common \emph{nested} design. Specifically, suppose the initial two-arm study was designed to include only units with $A_Y = A_M$ (e.g., a subset of students who receive both extra time and the separate session from the NAEP data), and a subsequent study added units with $A_Y \neq A_M$ (e.g., the original NAEP data with all the combinations of extra time and the separate session). Given that the four-arm data includes both the two-arm sample and the additional units (i.e., two-arm data are nested within four-arm data), \eqref{eq-falsification mean} simplifies to:
\begin{align*}
H_{0(iii)}: 
    \EXP[ Y_{[4]}^{1,a_M} - Y_{[4]}^{0,a_M} \cond A_{Y,[4]}=A_{M,[4]}] = \EXP [Y_{[2]}^{1,a_M} - Y_{[2]}^{0,a_M}] \ ,
    \numeq 
    \label{eq-falsification mean 2} \\
    H_{0(iv)}: 
    \EXP[ Y_{[4]}^{a_Y,1} - Y_{[4]}^{a_Y,0} \cond A_{Y,[4]}=A_{M,[4]}] = \EXP [Y_{[2]}^{a_Y,1} - Y_{[2]}^{a_Y,0}] \ .
    \numeq 
    \label{eq-falsification mean 3}
\end{align*} 
A violation of \eqref{eq-falsification mean 2} provides evidence against \ref{assumption-2arm-isolation}-(i) or \ref{assumption-2arm-dismissible}-(i), provided the other identifying assumptions hold. Similarly, a violation of \eqref{eq-falsification mean 3} provides evidence against \ref{assumption-2arm-isolation}-(ii) or \ref{assumption-2arm-dismissible}-(ii), provided the other identifying assumptions hold. Within each null hypothesis, the left-hand side is estimated based on the four-arm EIF estimators using the full four-arm sample, but it applies different weights in the propensity and outcome models to target the two-arm equivalent subgroup. Conceptually, this is similar to estimating the average treatment effect on the treated in binary treatment settings; although the estimand is defined only among the treated, the estimator utilizes all observations, but with different weights. In contrast, the right-hand side is estimated directly using the two-arm estimators discussed in Section \ref{sec:estimation_two}. 

We can formally test whether these two estimates significantly differ by using a Wald test. If this Wald test is significant, it indicates that the key identifying assumptions required for the separable effects of the two-arm design are likely invalid. Detailed derivations of the EIF-based estimator for the left-hand side, as well as the associated test statistic, are provided in Supplementary Appendix \ref{app:third estimator}.

\section{Simulation Study}\label{sec:simu}

\subsection{Designs and Evaluation}

We conducted simulation studies to evaluate the finite-sample performance of the two- and four-arm design estimators. The data-generation process followed the four-arm design structure. First, for each unit, we generated a 5-dimensional vector of baseline covariates $X=(X_1,\ldots,X_5)$ where each $X_{j}$ was independently drawn from a Bernoulli distribution with a success probability of $0.5$. Next, the treatment assignment $A_M$ was generated as $A_M \sim \text{Ber} \big( \text{expit}(-0.5 + 0.1\sum_{j=1}^{5} X_j) \big)$, and $A_Y$ was generated independently of $A_M$ based on one of the following two models (unbalanced and balanced):
\begin{align*}
    & \text{Model 1}: \
    &&
    A_Y \sim \text{Ber} \bigg( \text{expit} \bigg( -0.5 + 0.1\sum_{j=1}^{5} X_j \bigg) \bigg)
    \ , 
    \\
    & \text{Model 2}: \
    &&
    A_Y \sim \text{Ber} (0.5)
    \ .
\end{align*}
We then generated a 2-dimensional mediator vector $M^{a_M}=(M_1^{a_M},M_2^{a_M})$. The components $M_1^{a_M}$ and $M_2^{a_M}$ were generated independently as follows:
\begin{align*}
    &
    M_k^{a_M} \sim N \bigg(
    0.1 a_M + 0.5 \sum_{j=1}^{5} (X_j-0.5)
    ,
    0.5^2
    \bigg) \ , \quad k \in \{1,2\} \ .
\end{align*}
Finally, the potential outcome $Y^{a_Y,a_M}$ was generated from
\begin{align*}
    &
    Y^{a_Y,a_M}
    \sim 
    N
    \bigg(  
    t(X) a_Y
    +
    M_1^{a_M}
    +
    M_2^{a_M}
    +
    m(X)
    , 0.5^2
    \bigg)
    \ , 
    \\
    &
    t(X) = 2
    +
    0.25 \sum_{j=1}^{3} (X_j-0.5)
    -
    0.1 \sum_{j=4}^{5} (X_j-0.5)  \ , \\
    &
    m(X) = 0.2 \sum_{j=1}^{3} (X_j-0.5)
    +
    0.6 \sum_{j=4}^{5} (X_j-0.5) \ .
\end{align*}
The function $t(X)$ represents the treatment effect model, and $m(X)$ represents the outcome main effect model. The observed mediator and outcome were given by $M = M^{A_M}$ and $Y = Y^{A_Y,A_M}$. We varied the total sample size with $N \in \{1000,2000,4000,8000\}$. From the four-arm dataset of size $N$, we constructed the two-arm design data by retaining only observations for which $A_Y = A_M$. 

Since the four-arm and two-arm datasets in the simulation are constructed in the nested design introduced in Section \ref{sec: falsification SEE}, we use the subscripts $_{[4]}$ and $_{[2]}$ to distinguish between the four-arm and two-arm study populations, respectively. We denote the separable effects over the four-arm and two-arm design populations using the same subscripts, e.g., $\text{SDE}_{[4]}(a_M) = \EXP [ Y_{[4]}^{1,a_M} - Y_{[4]}^{0,a_M} ]$ and $\text{SDE}_{[2]}(a_M) = \EXP [ Y_{[2]}^{1,a_M} - Y_{[2]}^{0,a_M} ]$. In addition, because the data-generating process satisfies the transportability and nested design conditions described in Section \ref{sec: falsification SEE}, the separable effects in the two-arm population can be expressed as in Equations \eqref{eq-falsification mean 2} and \eqref{eq-falsification mean 3}. Therefore, we compared the following three estimators for estimating $\text{SDE}_{[\star]}(a_M)$ and $\text{SIE}_{[\star]}(a_Y)$ for $\star \in \{2,4\}$:
\begin{itemize}
\item ($\widetilde{\text{SDE}}_{[4]}$, $\widetilde{\text{SIE}}_{[4]}$) The four-arm design estimators for $\text{SDE}_{[4]}$ and $\text{SIE}_{[4]}$  based on $O_{[4]}$, defined in \eqref{eq-4arm estimator};
\item ($\widehat{\text{SDE}}_{[2]}$, $\widehat{\text{SIE}}_{[2]}$) The two-arm design estimators for $\text{SDE}_{[2]}$ and $\text{SIE}_{[2]}$ based on $O_{[2]}$, defined in \eqref{eq-2arm estimator};
\item ($\widetilde{\text{SDE}}_{[2]}$, $\widetilde{\text{SIE}}_{[2]}$)  The four-arm design estimators  for $\text{SDE}_{[2]}$ and $\text{SIE}_{[2]}$ based on $O_{[4]}$, defined in \eqref{eq-theta estimator} in Supplementary Appendix \ref{app:third estimator}. 
\end{itemize}
Here, the tilde $(\ \widetilde{} \ )$ and hat $(\ \widehat{}\ )$ notation indicate that the estimator is based on four-arm and two-arm data, respectively, while the subscript represents the population over which the target estimand is defined.

For estimating the nuisance functions, we used the \texttt{ranger} function from the R package \texttt{ranger} \parencite{ranger}. In the \texttt{ranger} function, we set the number of candidate variables for splitting at each node to 3 and varied the number of trees across 500, 1,000, and 1,500. We then combined the predictions from these random forests using a super learner algorithm \parencite{superlearer}. For the two-arm design estimator, we additionally trained random forests either by including treatment as a regressor in the estimation (S-learner) or by splitting the training data by treatment levels (T-learner), and combined their predictions by averaging; see Supplementary Appendix \ref{app:second estimator} for details. For all the estimators, we further apply the median-adjustment procedure described in Section \ref{sec:estimation_four} using three sample splits.

We evaluated the performance of each estimator relative to its specific population and the true target separable effect. We ran the simulation 1,000 times for each total sample size condition, $N \in \{1000,2000,4000,8000\}$. Specifically, for each estimator, we report the empirical bias, root mean squared error (RMSE), and coverage probability of the corresponding 95\% confidence intervals across 1,000 replications, i.e.,
\begin{align*}
    &
    \text{Bias}
    =
    \frac{1}{1000}
    \sum_{\ell=1}^{1000} 
    \big(
    \widehat{\tau}_{\ell} - \tau^*
    \big)
    \ ,
    \\
    &
    \text{RMSE}
    =
    \bigg\{
    \frac{1}{1000}
    \sum_{\ell=1}^{1000} 
    \big(
    \widehat{\tau}_{\ell} - \tau^*
    \big)^2
    \bigg\}^{1/2}
    \ ,
    \\
    &
    \text{Coverage}
    = 
    \frac{1}{1000}
    \sum_{\ell=1}^{1000} 
    \ind \big\{ \tau^* \in  \widehat{\text{CI}}_{\ell}\big\}
    \ ,
\end{align*}
where $\tau^*$ denotes the true target separable effect, $\widehat{\tau}_{\ell}$ is the estimate obtained in the $\ell$-th simulation replication, and $\widehat{\text{CI}}_{\ell}$ is the corresponding 95\% confidence interval.

\subsection{Simulation Results}

Table \ref{tab:tab1}  summarizes the results for the SDE. Across different sample size conditions, the estimators exhibit consistency for their respective target SDEs. Moreover, the RMSE values decrease at approximately the expected $\sqrt{N}$ rate, and the empirical coverage probabilities are close to the nominal 95\% level. Overall, the simulation results support the theoretical properties established for the proposed estimators. When comparing $\widehat{\text{SDE}}_{[2]}$ and $\widetilde{\text{SDE}}_{[2]}$, we observe that $\widehat{\text{SDE}}_{[2]}$ often achieves smaller RMSE values, despite relying on the subset of observations with $A_M = A_Y$, while $\widetilde{\text{SDE}}_{[2]}$ uses the entire sample of size $N$. 

\begin{table}[!htp]
\renewcommand{\arraystretch}{1.4} \centering
\setlength{\tabcolsep}{4pt}
\caption{Simulation results for the separable direct effect (SDE) estimators.}\label{tab:tab1}
\small
\begin{tabular}{c|c|ccc|ccc|ccc}
\hline
\multirow{2}{*}{$A_Y$ Model} & \multirow{2}{*}{$N$} & \multicolumn{3}{c|}{$\widetilde{\text{SDE}}_{[4]}(a_M=1)$}  & \multicolumn{3}{c|}{$\widehat{\text{SDE}}_{[2]}(a_M=1)$}     & \multicolumn{3}{c}{$\widetilde{\text{SDE}}_{[2]}(a_M=1)$}  \\ %\cline{3-11} 
                       &                      & \multicolumn{1}{c}{Bias}  & \multicolumn{1}{c}{RMSE}   & Coverage & \multicolumn{1}{c}{Bias}   & \multicolumn{1}{c}{RMSE}   & Coverage & \multicolumn{1}{c}{Bias}  & \multicolumn{1}{c}{RMSE}   & Coverage \\ \hline
\multirow{4}{*}{1}     & 1000                 & \multicolumn{1}{c}{0.854} & \multicolumn{1}{c}{9.583}  & 0.973    & \multicolumn{1}{c}{-3.553} & \multicolumn{1}{c}{10.336} & 0.931    & \multicolumn{1}{c}{0.345} & \multicolumn{1}{c}{9.770}  & 0.967    \\ %\cline{2-11} 
                       & 2000                 & \multicolumn{1}{c}{0.402} & \multicolumn{1}{c}{5.998}  & 0.963    & \multicolumn{1}{c}{-0.400} & \multicolumn{1}{c}{5.856}  & 0.946    & \multicolumn{1}{c}{0.273} & \multicolumn{1}{c}{6.076}  & 0.961    \\ %\cline{2-11} 
                       & 4000                 & \multicolumn{1}{c}{0.252} & \multicolumn{1}{c}{4.031}  & 0.962    & \multicolumn{1}{c}{-0.067} & \multicolumn{1}{c}{3.894}  & 0.950    & \multicolumn{1}{c}{0.131} & \multicolumn{1}{c}{4.063}  & 0.960    \\ %\cline{2-11} 
                       & 8000                 & \multicolumn{1}{c}{0.232} & \multicolumn{1}{c}{2.667}  & 0.971    & \multicolumn{1}{c}{0.312}  & \multicolumn{1}{c}{2.717}  & 0.940    & \multicolumn{1}{c}{0.182} & \multicolumn{1}{c}{2.667}  & 0.968    \\ \hline
\multirow{4}{*}{2}     & 1000                 & \multicolumn{1}{c}{2.774} & \multicolumn{1}{c}{17.621} & 0.980    & \multicolumn{1}{c}{0.758}  & \multicolumn{1}{c}{18.636} & 0.971    & \multicolumn{1}{c}{2.207} & \multicolumn{1}{c}{17.791} & 0.976    \\ %\cline{2-11} 
                       & 2000                 & \multicolumn{1}{c}{0.722} & \multicolumn{1}{c}{9.557}  & 0.972    & \multicolumn{1}{c}{1.747}  & \multicolumn{1}{c}{9.743}  & 0.971    & \multicolumn{1}{c}{0.523} & \multicolumn{1}{c}{9.638}  & 0.978    \\ %\cline{2-11} 
                       & 4000                 & \multicolumn{1}{c}{0.583} & \multicolumn{1}{c}{6.005}  & 0.963    & \multicolumn{1}{c}{1.334}  & \multicolumn{1}{c}{5.632}  & 0.955    & \multicolumn{1}{c}{0.511} & \multicolumn{1}{c}{6.036}  & 0.959    \\ %\cline{2-11} 
                       & 8000                 & \multicolumn{1}{c}{0.421} & \multicolumn{1}{c}{4.035}  & 0.957    & \multicolumn{1}{c}{1.141}  & \multicolumn{1}{c}{3.557}  & 0.956    & \multicolumn{1}{c}{0.393} & \multicolumn{1}{c}{4.037}  & 0.955    \\ \hline
\end{tabular}
\vspace{-0.1in}  
\begin{flushleft}
\footnotesize  NOTE: Bias and RMSE are multiplied by 100.
\end{flushleft}
\end{table}

Table \ref{tab:tab2} summarizes the results for the SIE. We find that the theoretical properties are guaranteed for the proposed estimators, similar to the SDE results. Across all sample size conditions, the estimators are consistent with respect to their target SIEs. The RMSE values decrease at the expected $\sqrt{N}$ rate, and the empirical coverage probabilities approximate the nominal 95\% level. Regarding the results from $\widehat{\text{SIE}}_{[2]}$ and $\widetilde{\text{SIE}}_{[2]}$, we find $\widehat{\text{SIE}}_{[2]}$ uniformly achieves smaller RMSE than $\widetilde{\text{SIE}}_{[2]}$ across all $N$ under both $A_Y$ models.

\begin{table}[!htp]
\renewcommand{\arraystretch}{1.4} \centering
\setlength{\tabcolsep}{4pt}
\caption{Simulation results for the separable indirect effect (SIE) estimators.}\label{tab:tab2}
\small
\begin{tabular}{c|c|ccc|ccc|ccc}
\hline
\multirow{2}{*}{$A_Y$ Model} & \multirow{2}{*}{$N$} & \multicolumn{3}{c|}{$\widetilde{\text{SIE}}_{[4]}(a_Y=1)$}  & \multicolumn{3}{c|}{$\widehat{\text{SIE}}_{[2]}(a_Y=1)$}     & \multicolumn{3}{c}{$\widetilde{\text{SIE}}_{[2]}(a_Y=1)$}  \\ %\cline{3-11} 
                       &                      & \multicolumn{1}{c}{Bias}  & \multicolumn{1}{c}{RMSE}   & Coverage & \multicolumn{1}{c}{Bias}   & \multicolumn{1}{c}{RMSE}   & Coverage & \multicolumn{1}{c}{Bias}  & \multicolumn{1}{c}{RMSE}   & Coverage \\ \hline
\multirow{4}{*}{1}     & 1000                 & \multicolumn{1}{c}{1.644} & \multicolumn{1}{c}{9.319}  & 0.968    & \multicolumn{1}{c}{2.825}  & \multicolumn{1}{c}{7.074}  & 0.987    & \multicolumn{1}{c}{1.272} & \multicolumn{1}{c}{9.550}  & 0.969    \\ %\cline{2-11} 
                       & 2000                 & \multicolumn{1}{c}{0.898} & \multicolumn{1}{c}{6.162}  & 0.961    & \multicolumn{1}{c}{0.829}  & \multicolumn{1}{c}{4.645}  & 0.982    & \multicolumn{1}{c}{0.748} & \multicolumn{1}{c}{6.254}  & 0.956    \\ %\cline{2-11} 
                       & 4000                 & \multicolumn{1}{c}{0.346} & \multicolumn{1}{c}{4.031}  & 0.960    & \multicolumn{1}{c}{0.163}  & \multicolumn{1}{c}{2.992}  & 0.982    & \multicolumn{1}{c}{0.266} & \multicolumn{1}{c}{4.135}  & 0.958    \\ %\cline{2-11} 
                       & 8000                 & \multicolumn{1}{c}{0.103} & \multicolumn{1}{c}{2.802}  & 0.955    & \multicolumn{1}{c}{-0.097} & \multicolumn{1}{c}{2.052}  & 0.972    & \multicolumn{1}{c}{0.060} & \multicolumn{1}{c}{2.826}  & 0.947    \\ \hline
\multirow{4}{*}{2}     & 1000                 & \multicolumn{1}{c}{5.214} & \multicolumn{1}{c}{12.742} & 0.978    & \multicolumn{1}{c}{7.928}  & \multicolumn{1}{c}{12.189} & 0.976    & \multicolumn{1}{c}{4.674} & \multicolumn{1}{c}{13.096} & 0.980    \\ %\cline{2-11} 
                       & 2000                 & \multicolumn{1}{c}{2.180} & \multicolumn{1}{c}{7.545}  & 0.960    & \multicolumn{1}{c}{2.295}  & \multicolumn{1}{c}{5.548}  & 0.980    & \multicolumn{1}{c}{1.908} & \multicolumn{1}{c}{7.514}  & 0.967    \\ %\cline{2-11} 
                       & 4000                 & \multicolumn{1}{c}{0.588} & \multicolumn{1}{c}{4.795}  & 0.960    & \multicolumn{1}{c}{0.734}  & \multicolumn{1}{c}{3.377}  & 0.979    & \multicolumn{1}{c}{0.444} & \multicolumn{1}{c}{4.771}  & 0.965    \\ %\cline{2-11} 
                       & 8000                 & \multicolumn{1}{c}{0.317} & \multicolumn{1}{c}{3.200}  & 0.967    & \multicolumn{1}{c}{0.109}  & \multicolumn{1}{c}{2.290}  & 0.970    & \multicolumn{1}{c}{0.245} & \multicolumn{1}{c}{3.249}  & 0.965    \\ \hline
\end{tabular}
\vspace{-0.1in} 
\begin{flushleft}
\footnotesize  NOTE: Bias and RMSE are multiplied by 100.
\end{flushleft}
\end{table}

\section{Empirical Example: ETA}\label{sec:ETAdata}

\subsection{Data and Variables}

For our real-data application, we used the 2017 NAEP Grade 8 process data for mathematics to study separable effects of ETA. The 2017 NAEP digital assessment for mathematics was administered using multiple test forms, where each form consists of two blocks. Within each test block, students were permitted to revisit and change their answers. In particular, we used process data from a single block that contains 15 items, including multiple choice items, short constructed items, and extended constructed response items. The items are publicly available at: \url{https://www.nationsreportcard.gov/nqt/searchquestions}. 

Our target treatment $A$ is ETA, which is decomposed into two binary components: extra time ($A_M \in \{\text{No=}0,\text{Yes=}1\}$) and a separate session ($A_Y \in \{\text{No=}0,\text{Yes=}1\}$). The extra time feature allows test-takers to use an additional 60 minutes beyond the standard 30 minutes for a given block. The separate session provides a different, often quieter or less distracting, environment rather than the standard testing setting. While extra time comes with a separate session in some testing conditions, the NAEP assessment design provided extra time both with and without a separate session. This design produced four-arm data: $(a_M,a_Y) \in \{(0,0),(0,1),(1,0),(1,1)\}$. For comparison, we also constructed two-arm data that contains only $(a_M,a_Y) \in \{(0,0),(1,1)\}$.

Our target sample focused on students eligible for ETA, i.e., students with disabilities and English language learners. From this sample, we excluded students who spent less than two minutes on the test to mitigate the influence of extremely non-effortful behavior. The final analytic sample for the four-arm data included 4,130 students, where 67\% were students with disabilities and 33\% were English language learners without disabilities. The corresponding two-arm data included 2,130 students. Note that numbers are unweighted and rounded to the nearest ten in accordance with NAEP's public reporting policy. 

The outcome of interest $Y$ is students' final test score, ranging from 0 to 25 points on the 15-item math test block. The mediator of interest $M_k$ is students' item access count (i.e., the number of items accessed) by a certain time point $k$ (e.g., the standard testing time of 30 minutes), and we used the set $\{M_{25}, M_{30}\}$. We constructed this mediator variable from the raw, unstructured process data by identifying whether a student exhibited a ``START'' action and spent more than three seconds on an item by the given time point.\footnote{A three-second threshold is used because shorter durations likely indicate a lack of effort.} We also used a set of pre-treatment covariates $X$, including student sex, race/ethnicity,  disability status, English learner status, free lunch status, parental education, and a proxy for the current math course level. See Supplementary Appendix \ref{app:descri} for more details on the variables used in NAEP data analysis, and also see Figure \ref{fig:DAG__ETA} for the assumed causal mechanism of the ETA. 

Following the methods proposed in Sections \ref{sec:sep_four} and \ref{sec:sep_two}, we applied our estimators to estimate the SDE and SIE of ETA. Specifically, we used the \texttt{ranger} function from the R package \texttt{ranger} \parencite{ranger} to fit random forests for the nuisance models. In the \texttt{ranger} function, we set the number of candidate variables for splitting at each node to 3 and varied the number of trees across 500, 1,000, and 1,500. We then combined the predictions from these random forests using a super learning algorithm \parencite{superlearer}. For the two-arm estimators, we additionally trained random forests using both S-learner and T-learner strategies, and combined their predictions by averaging as done in the simulations. Lastly, we conducted two distinct falsification tests for the key identification assumptions required for the two-arm design, as discussed in Section \ref{sec:falsification}. 

We remark that we did not apply sampling weights or jackknife replicate weights from the 2017 NAEP data to prioritize demonstrating our proposed methods. Therefore, our results are specific to our analytic samples and cannot be generalized to the broader population. \raggedbottom

\subsection{Results of Effect Estimates}

Table \ref{tab:freq} provides the contingency table between extra time and separate session statuses among students eligible for ETA. The majority of students received neither extra time nor a separate session. Among those who received extra time, 88\% took the NAEP math test in the standard testing environment, while 12\% utilized a separate session. 

\begin{table}[hbt!]
    \centering  
    \begin{threeparttable}
    \caption{Frequency of students who received extra time and separate sessions}\label{tab:freq}
\begin{tabular}{c C{1.5cm}  |R{3cm}R{3cm}R{3cm}}
\hline      
     &  & \multicolumn{3}{c}{Separate Session ($A_Y$)}  \\ 
    & & 0 & 1  & Total  \\  
\hline 
\multirow{3}{*}{Extra Time ($A_M$)} & 0 & 1,880 & 160 & 2,030 \\ 
&1 &  1,850 & 250 & 2,100 \\    
& Total & 3,730 & 410 & 4,130 \\    
\hline       
\end{tabular}
\vspace{1mm}
  \begin{tablenotes}[para,flushleft]
    \footnotesize NOTE: Numbers are unweighted and rounded to the nearest ten, and details may not sum to a total due to rounding. \\
    SOURCE: U.S. Department of Education, National Center for Education Statistics (NCES), 2017 NAEP Grade 8 Mathematics Process Data, Student Features Data File Partial Form and Response Data File.
  \end{tablenotes}
  \end{threeparttable}
\end{table}

Table \ref{tab:NEAP_effect} summarizes the estimates of the separable effects using (i) the four-arm estimators (i.e., $\widetilde{\text{SIE}}_{[4]}(a_Y)$, $\widetilde{\text{SDE}}_{[4]}(a_M)$), (ii) the two-arm estimators (i.e., $\widehat{\text{SIE}}_{[2]}(a_Y)$, $\widehat{\text{SDE}}_{[2]}(a_M)$), and (iii) the four-arm estimators for students with consistent treatment levels $A_Y=A_M$ (i.e., $\widetilde{\text{SIE}}_{[2]}(a_Y)$, $\widetilde{\text{SDE}}_{[2]}(a_M)$). Note that the two-arm estimators target the same population as the four-arm estimators for $A_Y=A_M$ in our NAEP data. Here, SIE($a_Y=1$) represents the average effect of extra time when a separate session is offered, while SIE($a_Y=0$) represents the average effect of extra time when the separate session is not offered. SDE($a_M=1$) denotes the average effect of a separate session when extra time is provided, whereas SDE($a_M=0$) denotes the average effect of a separate session in the absence of extra time. 
Across the different estimators, we found that only the SDE($a_M=0$) estimates are statistically significant and negative, while the other estimates are not significant. Since testing accommodations are designed to remove barriers in testing rather than strictly increase test scores, non-significant results are theoretically plausible. However, the finding that SDE($a_M=0$) has a negative effect contradicts the intended purpose of providing a less distracting environment. Nonetheless, this finding aligns with prior literature that reports neutral or even negative evidence on separate sessions. For example, in a randomized trial, \textcite{weis2020separate} found that students with disabilities performed worse in a separate room than in a standard group setting; that is, separate room testing did not effectively mitigate barriers for students with disabilities. 

\begin{table}[hbt!]
    \centering   
    \begin{threeparttable}
\setlength{\tabcolsep}{2pt}    
    \caption{Point estimates (Est.), standard errors (SE), and 95\% confidence intervals (CI) of the separable effects}\label{tab:NEAP_effect}
{\small \begin{tabular}{c|ccc|ccc|ccc}
\hline
Estimator    & \multicolumn{3}{c|}{Four-Arm} & \multicolumn{3}{c|}{Two-Arm} & \multicolumn{3}{c}{Four-Arm for $A_Y = A_M$} \\  
Effect    & Est. & SE & CI & Est. & SE & CI & Est. & SE & CI \\  
\hline 
 SIE($a_Y=1$) & -0.046 & 0.411 & (-0.851,\,0.759)  & \ 0.134  & 0.183 & (-0.226,\,0.493) & -0.060 & 0.415 & (-0.872,\,0.753) \\ 
 SIE($a_Y=0$) & -0.141 & 0.117 & (-0.371,\,0.089) & -0.218 & 0.138 & (-0.488,\,0.053) & -0.177 & 0.118 & (-0.408,\,0.054) \\ 
 SDE($a_M=1$) & -0.398 & 0.352 & (-1.088,\,0.293) & -0.318 & 0.351 & (-1.006,\,0.369) & -0.387 & 0.357 & (-1.087,\,0.313) \\
 SDE($a_M=0$) & -0.489 & 0.242 & (-0.962, -0.015) & -0.674 & 0.311 & (-1.284, -0.064)  & -0.503 & 0.239 & (-0.972, -0.035) \\ 
\hline       
\end{tabular}}
\vspace{1mm}
  \begin{tablenotes}[para,flushleft]
    \footnotesize NOTE: SDE = Separable direct effect; SIE = Separable indirect effect.  \\
    SOURCE: U.S. Department of Education, National Center for Education Statistics (NCES), 2017 NAEP Grade 8 Mathematics Process Data, Student Features Data File Partial Form and Response Data File. 
  \end{tablenotes}
  \end{threeparttable}
\end{table}

\subsection{Results of Falsification Tests}

Table \ref{tab:falsi} summarizes the results of our direct tests to falsify the dismissible conditions, as discussed in Section \ref{sec:Fal_reg}. We fitted linear regression models to test $H_{0(i)}$ and $H_{0(ii)}$ using the four-arm data. To test $H_{0(i)}$, we applied these regression models to the number of items accessed by 25 minutes ($M_{25}$) and 30 minutes ($M_{30}$). The resulting confidence intervals included zero, indicating no strong evidence to support a violation of the first dismissible condition. We also tested $H_{0(ii)}$ using the set $\{M_{25}, M_{30}\}$ as additional regressors. Similarly, the corresponding confidence intervals included zero, suggesting no strong evidence of a violation of the second dismissible condition. We note that these tests can also be used to falsify Assumption \ref{assumption-2arm-isolation}, provided certain assumptions are met; see Section \ref{sec:Fal_reg} for more details. 

\begin{table}[hbt!]
    \centering  
    \begin{threeparttable}
    \caption{Direct falsification tests for the dismissible conditions}\label{tab:falsi}       
\begin{tabular}{L{4.5cm}  |C{2.5cm}C{3.5cm}C{3.5cm}}
\hline      
    & Estimate & Standard Error & Confidence Interval \\  
\hline 
Test for $H_{0(i)}$ on $M_{25}$ & -0.092 & 0.109 & (-0.305, 0.121) \\ 
Test for $H_{0(i)}$ on $M_{30}$ & -0.040 & 0.084 & (-0.204, 0.123) \\ 
Test for $H_{0(ii)}$ & -0.118 & 0.107 & (-0.328, 0.091) \\  
\hline       
\end{tabular}
\vspace{1mm}
  \begin{tablenotes}[para,flushleft]
    \footnotesize 
    SOURCE: U.S. Department of Education, National Center for Education Statistics (NCES), 2017 NAEP Grade 8 Mathematics Process Data, Student Features Data File Partial Form and Response Data File.
  \end{tablenotes}
  \end{threeparttable}
\end{table}

Moreover, we conducted indirect tests comparing separable effect estimates obtained from the four-arm and two-arm data, as discussed in Section \ref{sec: falsification SEE}. Specifically, we tested whether these two estimates significantly differ by using a Wald test. As shown in Table \ref{tab:falsi2}, we found that none of the tests are significant, which indicates that the key identifying assumptions required for the separable effects in the two-arm design are not falsified. Overall, the results of the two distinct falsification tests increase the credibility of the effect estimates obtained from the two-arm estimators.

\begin{table}[hbt!]
    \centering  
    \begin{threeparttable}
    \caption{Indirect falsification tests}\label{tab:falsi2}       
\begin{tabular}{C{4.5cm}  |C{2.5cm}C{3.5cm}C{3.5cm}}
\hline      
    & Estimate & Standard Error & Test Statistic \\  
\hline 
SIE($a_Y=1$) & -0.204 & 0.430 & -0.473 \\ 
SIE($a_Y=0$) & \ 0.030 & 0.181 & \ 0.163 \\ 
SDE($a_M=1$) & -0.050 & 0.254 & -0.199 \\  
SDE($a_M=0$) & \ 0.169 & 0.351 & \ 0.480 \\  
\hline       
\end{tabular}
\vspace{1mm}
  \begin{tablenotes}[para,flushleft]
    \footnotesize 
    NOTE: SDE = Separable direct effect; SIE = Separable indirect effect. Wald test statistics are reported, and absolute values exceeding 1.96 indicate statistical significance ($p < .05$). \\
    SOURCE: U.S. Department of Education, National Center for Education Statistics (NCES), 2017 NAEP Grade 8 Mathematics Process Data, Student Features Data File Partial Form and Response Data File.
  \end{tablenotes}
  \end{threeparttable}
\end{table}

\section{Discussion and Conclusions}\label{sec:con}

This paper provides a general framework for estimating separable effects---SDE and SIE---in both four-arm and two-arm designs. Specifically, we detail the identification and estimation methods in each design, and propose new falsification tests to evaluate the validity of the underlying assumptions in the two-arm case. The proposed framework is the first to explicitly use four-arm data for separable effect estimation and to further expand its utility as a falsification tool for two-arm designs. Our simulation study reveals that the proposed estimators achieve the desired theoretical properties of consistency and asymptotic normality with respect to their target separable effects. Moreover, we demonstrate the effectiveness of our proposed methods by studying the separable effects of ETA on math scores using the NAEP data, where ETA is decomposed into extra time and a separate session. Since we did not find significant results on the proposed falsification tests, we are more confident in the validity of the empirical results obtained from the two-arm data. Regarding the estimates of the separable effects, only the effect of providing a separate session without extra time is significantly negative, while the other three separable effects are not statistically significant across the different estimators.

Our identification strategy in the two-arm design aligns primarily with the approach of Robins and colleagues \parencite{robins2010alternative, robins2022interventionist, Stensrud2021, Stensrud2022}, who use dismissible conditions (i.e., conditional independence relations in a hypothetical four-arm study) to avoid invoking both the cross-world independence assumption and direct interventions on the mediator. While these previous studies analyzed two-arm data, our framework bridges four-arm and two-arm designs and introduces two falsification tests for the dismissible conditions using four-arm data. Recently, \textcite{pitts2025addressing} conducted separable effects analyses following the treatment decomposition approach of \textcite{robins2010alternative} to address an extreme positivity violation. To identify these separable effects, they use the potential outcome that depends explicitly on the mediator, i.e., $Y(a,m)$, similar to \textcite{pearl2001direct}. As a result, the approach of \textcite{pitts2025addressing} can be viewed as a compromise between the treatment decomposition approach of \textcite{robins2010alternative} and the mediation framework of \textcite{pearl2001direct}. Importantly, although it does not depend on a cross-world assumption, this approach requires assuming the conditional independence assumption $Y(a,m) \indep M \cond (A = a, X)$ for all $m$. This assumption cannot be empirically falsified, even in future randomized experiments, because $Y(a,m)$ cannot be observed for all values of $m$ among individuals with $A = a$. This shares the same limitation as \citeauthor{pearl2001direct}'s mediation framework.

Furthermore, our separable effect results in the NAEP can inform the design and implementation of testing accommodations, which might otherwise be overlooked in prior studies treating ETA as a single intervention \parencite[e.g.,][]{lee2025evidence, suk2026causal}. As mentioned above, our NAEP data analysis finds a significantly negative effect of providing a separate session without extra time. Although the separate session accommodation is designed to provide a quiet and less distracted setting, it can still be a \textit{new} environment that accommodated students need time to adapt to. In addition, our results align with prior research \parencite{weis2020separate} revealing that students with disabilities performed worse in a separate room than in a standard testing setting. Given this prior work and our findings, educators and test administrators should provide extra time if students are accommodated with a separate session in order to mitigate its potential negative effects. In this way, we can achieve more informative policy implications via separable effect analysis, and these benefits will translate to other interventions analyzed under this framework.    

Finally, we acknowledge some limitations of this paper and offer suggestions for future research. First, we assume that all the confounders are measured in the observed data. Since there may be unmeasured confounders, future research would apply sensitivity analysis techniques, like those proposed by \textcite{Stensrud2021} and \textcite{pitts2025addressing}, or utilize causal methods robust to unmeasured confounding, such as proximal causal inference \parencite[e.g.,][]{park2024proximal}. Second, the proposed framework assumes that the treatment consists of exactly two components operating through a single mediating pathway. In practice, treatments may involve more than two components, and extending the framework to handle such cases (e.g., defining meaningful causal estimands; developing efficient estimation methods) remains an open challenge. Third, while we focus on binary components in this paper, adapting the identification and estimation strategies to multi-valued or continuous components would make the framework applicable to a broader set of interventions. Fourth, we focus on the effects of treatment components on the final outcome (e.g., the final test score), but their effects may be dynamic over time. Future research could extend our framework to time-varying outcome settings (e.g., cumulative test scores over time).   

\newpage

\printbibliography

\newpage  
\newpage

\appendix 

\setcounter{equation}{0}
\setcounter{figure}{0}
\setcounter{table}{0}
\setcounter{page}{1}
\setcounter{section}{0}
\renewcommand{\theequation}{S\arabic{equation}}
\renewcommand{\thefigure}{S\arabic{figure}}
\renewcommand{\thesection}{S\arabic{section}}
\renewcommand{\thetable}{S\arabic{table}}
\renewcommand{\thefigure}{S\arabic{figure}}

\section*{Supplementary Material}

\section{Regularity Conditions}\label{app:regu_cond}

Assumption \ref{assumption-reg-4arm} outlines the regularity conditions for the separable effect estimators in the four-arm design.

\begin{assumption} \label{assumption-reg-4arm}
    \begin{itemize}
        \item[(i)] There exist constants $c_{\nu}$, $c_\pi$, $c_Y$ such that $| \nu^*(a_Y,a_M,X) | \leq c_{\nu} $, $| \widehat{\nu}\LSS (a_Y,a_M,X) | \leq c_{\nu} $, $| 1/ \widehat{\pi}\LSS (a_Y,a_M,X) | \leq c_{\pi} $, $\EXP[Y^2 \cond A_Y=a_Y,A_M=a_M,X] \leq c_{Y}$ for all $k\in \{1,2\}$, $a_Y\in \{0,1\}$, $a_M\in \{0,1\}$, $X$. 
        \item[(ii)] For all $k\in \{1,2\}$, $\| \widehat{\delta}\LSS - \delta^* \|_{P,2} = o_P(N^{-1/4})$, where $\| g(O) \|_{P,2}^2 = \EXP[ g^2(o) ] $ is the $L_2(P)$-norm of $g$.
        
    \end{itemize}
\end{assumption}

Assumption \ref{assumption-reg-4arm}-(i) ensures that both the nuisance functions and their estimators are uniformly bounded, and the outcome has a finite conditional second moment. Assumption \ref{assumption-reg-4arm}-(ii) requires the estimators of the nuisance functions to converge to the true nuisance functions at $N^{1/4}$-rate. 

Assumption \ref{assumption-reg-2arm} outlines the regularity conditions for the separable effect estimators in the two-arm design.

\begin{assumption} \label{assumption-reg-2arm}
    \begin{itemize}
        \item[(i)] There exist constants $c_{\rho}$, $c_{\omega}$, $c_{\mu}$, $c_{\lambda}$, $c_Y$ such that 
        $\| 1/\rho^*(a,M,X) \| \leq c_{\rho}$, 
        $\| 1/\widehat{\rho}\LSS(a,M,X) \| \leq c_{\rho}$, 
        $\| 1/\omega^*(a,X) \| \leq c_{\omega}$, 
        $\| 1/\widehat{\omega}\LSS(a,X) \| \leq c_{\omega}$, 
        $\| \mu^*(a,M,X) \| \leq c_{\mu}$, 
        $\| \widehat{\mu}\LSS (a,M,X) \| \leq c_{\mu}$, 
        $\| \lambda^*(a,a',X) \| \leq c_{\lambda}$, 
        $\| \widehat{\lambda}\LSS (a,a',X) \| \leq c_{\lambda}$, $\EXP[Y^2 \cond A=a,X] \leq c_{Y}$       
        for all $k\in \{1,2\}$, $a\in \{0,1\}$, $a'\in \{0,1\}$, $M$, $X$.
        \item[(ii)] For all $k\in \{1,2\}$, $\| \widehat{\eta}\LSS - \eta^* \|_{P,2} = o_P(N^{-1/4})$.
    \end{itemize}
\end{assumption}
Assumption \ref{assumption-reg-2arm} parallels Assumption \ref{assumption-reg-4arm}, except that its conditions are related to the two-arm nuisance parameters $\eta^*$ rather than $\delta^*$. 

\section{Proof}

\subsection{Proof of Theorem \ref{thm-4arm}}

We present the results for the SDE; the same reasoning applies to derive the corresponding results for the SIE. 

Let $\Phi(O \con a_M, \delta) =\phi(O \con 1,a_M, \delta) - \phi(O \con 0,a_M, \delta)$ be the EIF for $\text{SDE}^*(a_M)$. Let $\EXP \LSS \{ g(O \con \widehat{\delta}\LSS ) \} = \int g(o \con \widehat{\delta}\LSS ) f^*(o) \, do $ be the expectation, treating $\widehat{\delta}\LSS$ as a fixed function.

The normalized difference between the estimator and the true parameter is decomposed as:
\begin{align}
&
\sqrt{N}
\big\{ 
    \widetilde{\text{SDE}}(a_M)
    -
    \text{SDE}^*(a_M)
\big\}
\nonumber
\\
&
=
\frac{1}{\sqrt{N}}
\sum_{k=1}^{K} \sum_{i \in \mathcal{I}^{(k)}} 
\big\{ 
    \Phi(O_i \con a_M, \widehat{\delta}\LSS)
    -
    \text{SDE}^*(a_M)
    \big\} 
\nonumber
\\
&
=
\frac{1}{\sqrt{N}}
\sum_{i=1}^{N}
\big\{ 
    \Phi(O_i \con a_M, \delta^*)
    -
    \text{SDE}^*(a_M)
    \big\} 
    \nonumber
    \\
    &
    \qquad +
    \frac{1}{\sqrt{N}}
\sum_{k=1}^{K} \sum_{i \in \mathcal{I}^{(k)}} 
\big[
    \EXP\LSS \{ \Phi(O_i \con a_M, \widehat{\delta}\LSS) \}
    -
    \EXP\LSS \{ \Phi(O_i \con a_M, \delta^*) \}
    \big] 
    \label{term-3}
    \\
    &
    \qquad  +
    \frac{1}{\sqrt{N}}
\sum_{k=1}^{K} \sum_{i \in \mathcal{I}^{(k)}} 
\left[ 
\begin{array}{l}
    \Phi(O_i \con a_M, \widehat{\delta}\LSS)
    -
    \EXP\LSS \{ \Phi(O_i \con a_M, \widehat{\delta}\LSS) \} 
    \\
    -
    \Phi(O_i \con a_M, \delta^*)
    +    
    \EXP\LSS \{ \Phi(O_i \con a_M, \delta^*) \}
\end{array}
    \right]\ .
    \label{term-2} 
\end{align}
In what follows, we will establish \eqref{term-3} and \eqref{term-2} are $o_P(1)$. Therefore, from the central limit theorem, we have 
\begin{align*}
&
    \sqrt{N}
\big\{ 
    \widetilde{\text{SDE}}(a_M)
    -
    \text{SDE}^*(a_M)
\big\}
\\
&
=
\frac{1}{\sqrt{N}}
\sum_{i=1}^{N}
\big\{ 
    \Phi(O_i \con a_M, \delta^*)
    -
    \text{SDE}^*(a_M)
    \big\} 
    +
    o_P(1)
    \\
    &
    \stackrel{D}{\rightarrow}
    N \Big( 0,\sigma_{\text{SDE}}^2(a_M) \Big) \ .
\end{align*}
This establishes the asymptotic normality.

It remains to show that \eqref{term-3} and \eqref{term-2} are $o_P(1)$. To show that \eqref{term-3} is $o_P(1)$, note that the bias term satisfies:
\begin{align*}
    & 
    \Big\|
    \EXP\LSS \{ \Phi(O \con a_M, \widehat{\delta}\LSS) \}
    -
    \EXP\LSS \{ \Phi(O \con a_M, \delta^*) \} 
    \Big\|
    \\
    &
    =
    \left\|
    \EXP \LSS
    \left[ 
    \begin{array}{l}
    \frac{ \ind(A_Y=1,A_M=a_M) \{ Y - \widehat{\nu}\LSS(1,a_M,X) \}
    }{ \widehat{\pi}\LSS(1,a_M,X) } 
    +
    \widehat{\nu}\LSS(1,a_M,X)
    - \nu^*(1,a_M,X)
    \\
    -
    \frac{ \ind(A_Y=0,A_M=a_M) \{ Y - \widehat{\nu}\LSS(0,a_M,X) \}
    }{ \widehat{\pi}\LSS(0,a_M,X) } 
    -
    \widehat{\nu}\LSS(0,a_M,X)     
    + \nu^*(1,a_M,X)
    \end{array}
    \right]  
    \right\|
    \\
    &
    =
    \left\|
    \EXP \LSS
    \left[ 
    \begin{array}{l}
    \big[ 
    \frac{ \pi^*(1,a_M,X) 
    }{ \widehat{\pi}\LSS(1,a_M,X) } 
    - 1 \big] 
    \{ \nu^*(1,a_M,X) - \widehat{\nu}\LSS(1,a_M,X) \} 
    \\
    -
    \big[ 
    \frac{ \pi^*(0,a_M,X) 
    }{ \widehat{\pi}\LSS(0,a_M,X) } 
    - 1 \big] 
    \{ \nu^*(0,a_M,X) - \widehat{\nu}\LSS(0,a_M,X) \} 
    \end{array}
    \right]  
    \right\|
    \\
    &
    \lesssim 
    \sum_{a=0}^{1}
    \| \widehat{\tau}\LSS(a,a_M,X) - \tau^*(a,a_M,X) 
    \|_{P,2}
    \| \widehat{\nu}\LSS(a,a_M,X) - \nu^*(a,a_M,X) 
    \|_{P,2} 
    \\
    &
    =
    o_P(N^{-1/2})
    \ .
    \numeq \label{eq-proof1}
\end{align*}
The inequality follows from the Cauchy-Schwarz inequality and Assumption \ref{assumption-reg-4arm}-(i), while the final line follows from the rate conditions in Assumption \ref{assumption-reg-4arm}-(ii). Thus:
\begin{align*}
    \eqref{term-3} 
    & = 
    \frac{1}{\sqrt{N}}
\sum_{k=1}^{K} \sum_{i \in \mathcal{I}^{(k)}} 
\big[
    \EXP\LSS \{ \Phi(O_i \con a_M, \widehat{\delta}\LSS) \}
    -
    \EXP\LSS \{ \Phi(O_i \con a_M, \delta^*) \}
    \big]  
    \\
    &
    \lesssim
    \sqrt{N} 
    \big[ \EXP\LSS \{ \Phi(O \con a_M, \widehat{\delta}\LSS) \}
    -
    \EXP\LSS \{ \Phi(O \con a_M, \delta^*) \}  \big]
    \\
    & \stackrel{\eqref{eq-proof1}}{=} \sqrt{N} o_P(N^{-1/2})
    \\
    &
    = o_P(1) \ .
\end{align*}

Next, we establish that \eqref{term-2} is $o_P(1)$ by showing $\mathbbm{G}_{k} =o_P(1)$ for each $k$, where
\begin{align*}
\mathbbm{G}_{k} = 
\frac{1}{\sqrt{N/2}}
 \sum_{i \in \mathcal{I}^{(k)}} 
\left[ 
\begin{array}{l}
    \Phi(O_i \con a_M, \widehat{\delta}\LSS)
    -
    \EXP\LSS \{ \Phi(O_i \con a_M, \widehat{\delta}\LSS) \} 
    \\
    -
    \Phi(O_i \con a_M, \delta^*)
    +    
    \EXP\LSS \{ \Phi(O_i \con a_M, \delta^*) \}
\end{array}
    \right] \ .
\end{align*}  

By construction, $\EXP\LSS[ \mathbbm{G}_{k} ] = 0$. Therefore, it suffices to show that $\VAR\LSS[ \mathbbm{G}_{k} ] = o_P(1)$. Specifically, we obtain
\begin{align*}
    &
    \VAR\LSS[ \mathbbm{G}_{k} ]
    \\
    & = 
    \VAR\LSS 
    \left[ 
\begin{array}{l}
    \Phi(O_i \con a_M, \widehat{\delta}\LSS)
    -
    \EXP\LSS \{ \Phi(O_i \con a_M, \widehat{\delta}\LSS) \} 
    \\
    -
    \Phi(O_i \con a_M, \delta^*)
    +    
    \EXP\LSS \{ \Phi(O_i \con a_M, \delta^*) \}
\end{array}
    \right]
    \\
    & = 
    \EXP \LSS 
    \left[ 
    \left\{
    \Phi(O_i \con a_M, \widehat{\delta}\LSS)
    -
    \Phi(O_i \con a_M, \delta^*) 
\right\}^2
    \right]
    \\
    & = 
    \EXP \LSS 
    \left[ 
    \left\{
    \begin{array}{l}
    \ind(A_Y=1,A_M=a_M) Y
    \big\{  \frac{1}{\widehat{\pi}\LSS(1,a_M,X)} - \frac{1}{\pi^*(1,a_M,X)} \big\}
    \\
    +
    \ind(A_Y=1,A_M=a_M) 
    \big\{ \frac{ \widehat{\nu}\LSS(1,a_M,X)}{ \widehat{\pi}\LSS(1,a_M,X)} - \frac{\nu^*(1,a_M,X)}{\pi^*(1,a_M,X)} \big\}
    \\
    -
    \widehat{\nu}\LSS(0,a_M,X) + \nu^*(0,a_M,X)
    \\
    -
    \ind(A_Y=0,A_M=a_M) Y
    \big\{  \frac{1}{\widehat{\pi}\LSS(0,a_M,X)} - \frac{1}{\pi^*(0,a_M,X)} \big\}
    \\
    -
    \ind(A_Y=0,A_M=a_M) 
    \big\{ \frac{ \widehat{\nu}\LSS(0,a_M,X)}{ \widehat{\pi}\LSS(0,a_M,X)} - \frac{\nu^*(0,a_M,X)}{\pi^*(0,a_M,X)} \big\}
    \\
    +
    \widehat{\nu}\LSS(0,a_M,X) - \nu^*(0,a_M,X)
    \end{array}
\right\}^2
    \right]
    \\
    &
    \lesssim 
    \sum_{a=0}^{1}
    \big[ 
    \big\| \widehat{\pi}\LSS(a,a_M,X) - \pi^*(a,a_M,X) \big\|_{P,2}^2
    +
    \big\| \widehat{\nu}\LSS(a,a_M,X) - \nu^*(a,a_M,X) \big\|_{P,2}^2
    \big] 
    \\
    &
    =
    o_P(1) 
    \ .
    \numeq \label{eq-proof3}
\end{align*}
The inequality follows from the Cauchy-Schwarz inequality and Assumption \ref{assumption-reg-4arm}-(i), while the final line follows from the rate conditions in Assumption \ref{assumption-reg-4arm}-(ii). 

Finally, we establish the consistency of the variance estimator. Note that $\widetilde{\sigma}_{\text{SDE}}^2(a_M) = \sum_{k=1}^{K} \widetilde{\sigma}_{k,\text{SDE}}^2(a_M)/K$ where
\begin{align*}
    &
    \widetilde{\sigma}_{k,\text{SDE}}^2(a_M)
    =
    \frac{1}{ N/K } 
    \sum_{i \in \mathcal{I}^{(k)} }
    \big\{ 
        \Phi(O_i \con a_M, \widehat{\delta}\LSS)
    -
    \widetilde{\text{SDE}}(a_M)
    \big\}^2  \ .
    \end{align*}
It suffices to show the consistency of $\widetilde{\sigma}_{k,\text{SDE}}^2(a_M)$, i.e., $\widetilde{\sigma}_{k,\text{SDE}}^2(a_M) - \sigma_{\text{SDE}}^2(a_M) = o_P(1)$. We can express this difference as:
\begin{align*}
    &
    \widetilde{\sigma}_{k,\text{SDE}}^2(a_M) - \sigma_{\text{SDE}}^2(a_M)
    \\
    &
    =
    \frac{1}{N/2}
    \sum_{i \in \mathcal{I}^{(k)}} 
    \big\{ 
        \Phi(O_i \con a_M, \widehat{\delta}\LSS)
    -
    \widetilde{\text{SDE}}(a_M)
    \big\}^2 - \sigma_{\text{SDE}}^2(a_M)
    \\
    &
    =
     \frac{1}{N/2}
    \sum_{i \in \mathcal{I}^{(k)}} 
    \Big[ 
    \big\{ 
        \Phi(O_i \con a_M, \widehat{\delta}\LSS)
    -
    \widetilde{\text{SDE}}(a_M)
    \big\}^2
    -
    \big\{ 
        \Phi(O_i \con a_M, \delta^*)
    -
    \text{SDE}^*(a_M)
    \big\}^2
    \Big] 
    \\
    & \qquad 
    +
    \frac{1}{N/2}
    \sum_{i \in \mathcal{I}^{(k)}} 
    \big\{ 
        \Phi(O_i \con a_M, \delta^*)
    -
    \text{SDE}^*(a_M)
    \big\}^2  - \sigma_{\text{SDE}}^2(a_M)
    \\
    &
    = 
    \frac{1}{N/2}
    \sum_{i \in \mathcal{I}^{(k)}} 
    \Big[ 
    \big\{ 
        \Phi(O_i \con a_M, \widehat{\delta}\LSS)
    -
    \widetilde{\text{SDE}}(a_M)
    \big\}^2
    -
    \big\{ 
        \Phi(O_i \con a_M, \delta^*)
    -
    \text{SDE}^*(a_M)
    \big\}^2
    \Big]
    + o_P(1) \ .
    \numeq
    \label{eq-proof2}
\end{align*}
The last line holds from the law of large numbers. 

Let $\widehat{\Delta}_i\LSS = \{ 
        \Phi(O_i \con a_M, \widehat{\delta}\LSS)
    -
    \widetilde{\text{SDE}}(a_M)
    \}
    -
    \{ 
        \Phi(O_i \con a_M, \delta^*)
    -
    \text{SDE}^*(a_M)
    \}$. Then, \eqref{eq-proof2} can be represented as 
\begin{align*}
    &
    \frac{1}{N/2}
    \sum_{i \in \mathcal{I}^{(k)}} 
    \Big[ 
    \big\{ 
        \Phi(O_i \con a_M, \widehat{\delta}\LSS)
    -
    \widetilde{\text{SDE}}(a_M)
    \big\}^2
    -
    \big\{ 
        \Phi(O_i \con a_M, \delta^*)
    -
    \text{SDE}^*(a_M)
    \big\}^2
    \Big]
    \\
    &
    =
    \frac{1}{N/2}
    \sum_{i \in \mathcal{I}^{(k)}} 
\widehat{\Delta}_i\LSS
    \times 
    \big[
\widehat{\Delta}_i\LSS
    +
    2
    \{ 
        \Phi(O_i \con a_M, \delta^*)
    -
    \text{SDE}^*(a_M)
    \}
    \big]
    \\
    &
    =
    \frac{1}{N/2}
    \sum_{i \in \mathcal{I}^{(k)}} 
\{ \widehat{\Delta}_i\LSS \}^2
+
    \frac{2}{N/2}
    \sum_{i \in \mathcal{I}^{(k)}} 
    \widehat{\Delta}_i\LSS 
    \{ 
        \Phi(O_i \con a_M, \delta^*)
    -
    \text{SDE}^*(a_M)
    \}  
    \ .
\end{align*}
 From the H\"older's inequality, we find the absolute value of \eqref{eq-proof2} is bounded by
 \begin{align*}
     &
     \big\| \eqref{eq-proof2} \big\|
     \\
     &
     \leq 
         \frac{1}{N/2}
    \sum_{i \in \mathcal{I}^{(k)}} 
\{ \widehat{\Delta}_i\LSS \}^2
+
2
\sqrt{
    \frac{1}{N/2}
    \sum_{i \in \mathcal{I}^{(k)}} 
    \{ \widehat{\Delta}_i\LSS \}^2
    }
    \underbrace{
    \sqrt{
    \frac{1}{N/2}
    \sum_{i \in \mathcal{I}^{(k)}} 
    \{ 
        \Phi(O_i \con a_M, \delta^*)
    -
    \text{SDE}^*(a_M)
    \}  ^2
    } }_{\sigma^2+o_P(1)=O_P(1)}
    \ .
 \end{align*} 
 Therefore, \eqref{eq-proof2} is $o_P(1)$ if $\frac{1}{N/2}
    \sum_{i \in \mathcal{I}^{(k)}} 
    \{ \widehat{\Delta}_i\LSS \}^2 = o_P(1)$. From some algebra, we find
    \begin{align*}
    &
    \frac{1}{N/2}
    \sum_{i \in \mathcal{I}^{(k)}} 
    \{ \widehat{\Delta}_i\LSS \}^2
    \\
    &
    =
    \frac{1}{N/2}
    \sum_{i \in \mathcal{I}^{(k)}} 
    \Big[ 
    \{ 
    \Phi(O_i \con a_M, \widehat{\delta}\LSS)
    -
    \widetilde{\text{SDE}}(a_M)
    \}
    -
    \{ 
        \Phi(O_i \con a_M, \delta^*)
    -
    \text{SDE}^*(a_M)
    \}
    \Big]^2
    \\
    &
    \lesssim 
    \frac{1}{N/2}
    \sum_{i \in \mathcal{I}^{(k)}}  
    \{ 
    \Phi(O_i \con a_M, \widehat{\delta}\LSS)
    -
        \Phi(O_i \con a_M, \delta^*)
    \}^2
    +
    \{ \widetilde{\text{SDE}}(a_M) - \text{SDE}^*(a_M) \}^2
    \\
    &
    =
    \EXP\LSS
    \big[ 
    \{ 
    \Phi(O_i \con a_M, \widehat{\delta}\LSS)
    -
        \Phi(O_i \con a_M, \delta^*)
    \}^2
    \big] 
    +
    o_P(1)
    \\
    &
    =
    o_P(1) \ .
    \end{align*}
The fourth line holds from $\widetilde{\text{SDE}}(a_M) - \text{SDE}^*(a_M) = o_P(1) $ and the law of large numbers. The last line holds from \eqref{eq-proof3}. This concludes the proof.

\subsection{Proof of Theorem \ref{thm-2arm}}

We present the results for the SDE with $a_M=1$; the same reasoning applies to derive the corresponding results for the SDE with $a_M=0$ and the SIE. 

Let $\Psi(O \con \delta) =\psi(O \con a_Y=1,a_M=1, \eta) - \psi(O \con a_Y=0,a_M=1, \eta)$ be the EIF for $\text{SDE}^* = \text{SDE}^*(a_M=1)$. Let $\EXP \LSS \{ g(O \con \widehat{\eta}\LSS ) \} = \int g(o \con \widehat{\eta}\LSS ) f^*(o) \, do $ be the expectation, treating $\widehat{\eta}\LSS$ as a fixed function.

The normalized difference between the estimator and the true parameter is decomposed as:
\begin{align}
&
\sqrt{N}
\big\{ 
    \widehat{\text{SDE}} 
    -
    \text{SDE}^*(a_M=1)
\big\}
\nonumber
\\
&
=
\frac{1}{\sqrt{N}}
\sum_{k=1}^{K} \sum_{i \in \mathcal{I}^{(k)}} 
\big\{ 
    \Psi(O_i \con \widehat{\delta}\LSS)
    -
    \text{SDE}^*
    \big\} 
\nonumber
\\
&
=
\frac{1}{\sqrt{N}}
\sum_{i=1}^{N}
\big\{ 
    \Psi(O_i \con   \delta^*)
    -
    \text{SDE}^*
    \big\} 
    \nonumber
    \\
    &
    \qquad +
    \frac{1}{\sqrt{N}}
\sum_{k=1}^{K} \sum_{i \in \mathcal{I}^{(k)}} 
\big[
    \EXP\LSS \{ \Psi(O_i \con \widehat{\eta}\LSS) \}
    -
    \EXP\LSS \{ \Psi(O_i \con \eta^*) \}
    \big] 
    \label{term-3-2arm}
    \\
    &
    \qquad  +
    \frac{1}{\sqrt{N}}
\sum_{k=1}^{K} \sum_{i \in \mathcal{I}^{(k)}} 
\left[ 
\begin{array}{l}
    \Psi(O_i \con a_M, \widehat{\eta}\LSS)
    -
    \EXP\LSS \{ \Psi(O_i \con  \widehat{\eta}\LSS) \} 
    \\
    -
    \Psi(O_i \con  \eta^*)
    +    
    \EXP\LSS \{ \Psi(O_i \con \eta^*) \}
\end{array}
    \right]\ .
    \label{term-2-2arm} 
\end{align}
In what follows, we will establish \eqref{term-3-2arm} and \eqref{term-2-2arm} are $o_P(1)$. Therefore, from the central limit theorem, we have 
\begin{align*}
&
    \sqrt{N}
\big\{ 
    \widehat{\text{SDE}} 
    -
    \text{SDE}^* 
\big\}
=
\frac{1}{\sqrt{N}}
\sum_{i=1}^{N}
\big\{ 
    \Phi(O_i \con \delta^*)
    -
    \text{SDE}^* 
    \big\} 
    +
    o_P(1)
    \stackrel{D}{\rightarrow}
    N \Big( 0,\sigma_{\text{SDE}}^2 \Big) \ .
\end{align*}
This establishes the asymptotic normality.

It remains to show that \eqref{term-2-2arm} and \eqref{term-3-2arm} are $o_P(1)$. To show that \eqref{term-3-2arm} is $o_P(1)$, note that the bias term satisfies: 
\begin{align*}
    & 
    \EXP\LSS \{ \Psi(O \con  \widehat{\eta}\LSS) \}
    -
    \EXP\LSS \{ \Psi(O \con  \eta^*) \} 
    \\
    &
    =
    \EXP \LSS
    \left[ 
    \begin{array}{l}
    \frac{ A }{ \widehat{\omega}\LSS(1,X) } 
    \frac{ \widehat{\rho}\LSS(1,M,X) }{ \widehat{\rho}\LSS(1,M,X) } 
    \{ Y - \widehat{\mu}\LSS(1,M,X) \}
    \\
    +
    \frac{ A }{ \widehat{\omega}\LSS(1,X) } 
    \{ \widehat{\mu}\LSS(1,M,X) - \widehat{\lambda}\LSS(1,1,X) \}
    \\
    -
    \frac{ 1-A }{ \widehat{\omega}\LSS(1,X) } 
    \frac{ \widehat{\rho}\LSS(1,M,X) }{ \widehat{\rho}\LSS(0,M,X) } 
    \{ Y - \widehat{\mu}\LSS(0,M,X) \}
    \\
    -
    \frac{ A }{ \widehat{\omega}\LSS(1,X) } 
    \{ \widehat{\mu}\LSS(0,M,X) - \widehat{\lambda}\LSS(0,1,X) \}
    \\
    + \lambda\LSS(1,1,X)
    - \lambda^*(1,1,X)
    \\
    - \lambda\LSS(0,1,X)
    + \lambda^*(0,1,X)
    \end{array}
    \right]  
    \\
    &
    = 
    \EXP \LSS
    \left[ 
    \begin{array}{l}
    \frac{ A }{ \widehat{\omega}\LSS(1,X) } 
    \{ Y - \widehat{\lambda}\LSS(1,1,X) \} 
    + \lambda\LSS(1,1,X)
    - \lambda^*(1,1,X)
    \\
    -
    \frac{ 1-A }{ \widehat{\omega}\LSS(1,X) } 
    \frac{ \widehat{\rho}\LSS(1,M,X) }{ \widehat{\rho}\LSS(0,M,X) } 
    \{ \mu^*(0,M,X) - \widehat{\mu}\LSS(0,M,X) \}
    \\
    -
    \frac{ A }{ \widehat{\omega}\LSS(1,X) } 
    \{ \widehat{\mu}\LSS(0,M,X) - \widehat{\lambda}\LSS(0,1,X) \}
    \\
    - \lambda\LSS(0,1,X)
    + \lambda^*(0,1,X)
    \end{array}
    \right]  
    \ . 
    \numeq \label{eq-proof0-2arm}
\end{align*}

Following the same logic as \eqref{eq-proof1} in the four-arm design, the first line inside the expectation can be straightforwardly shown to be $o_P(N^{-1/2})$, i.e.,
\begin{align}
        \EXP \LSS
    \Bigg[ 
    \frac{ A }{ \widehat{\omega}\LSS(1,X) } 
    \{ Y - \widehat{\lambda}\LSS(1,1,X) \} 
    + \lambda\LSS(1,1,X)
    - \lambda^*(1,1,X)
    \Bigg] 
    =
    o_P(N^{-1/2})
    \ .
    \label{eq-proof00-2arm} 
\end{align}

Therefore, it suffices to show that the remaining terms in \eqref{eq-proof0-2arm} are $o_P(N^{-1/2})$. We find that:
\begin{align*}
&
\EXP \LSS
    \left[ 
    \begin{array}{l}
    \frac{ 1-A }{ \widehat{\omega}\LSS(1,X) } 
    \frac{ \widehat{\rho}\LSS(1,M,X) }{ \widehat{\rho}\LSS(0,M,X) } 
    \{ \mu^*(0,M,X) - \widehat{\mu}\LSS(0,M,X) \}
    \\
    +
    \frac{ A }{ \widehat{\omega}\LSS(1,X) } 
    \{ \widehat{\mu}\LSS(0,M,X) - \widehat{\lambda}\LSS(0,1,X) \}
    \\
    + \lambda\LSS(0,1,X)
    - \lambda^*(0,1,X)
    \end{array}
    \right]  
    \\
    &
    =
        \EXP \LSS
    \left[ 
    \begin{array}{l}
    \frac{ \rho^*(0,M,X) }{ \widehat{\omega}\LSS(1,X) } 
    \frac{ \widehat{\rho}\LSS(1,M,X) }{ \widehat{\rho}\LSS(0,M,X) } 
    \{ \mu^*(0,M,X) - \widehat{\mu}\LSS(0,M,X) \}
    \\
    +
    \frac{ \rho^*(1,M,X) }{ \widehat{\omega}\LSS(1,X) } 
    \{ \widehat{\mu}\LSS(0,M,X) - \widehat{\lambda}\LSS(0,1,X) \}
    \\
    \lambda\LSS(0,1,X)
    - \lambda^*(0,1,X)
    \end{array}
    \right] 
    \numeq \label{eq-proof01-2arm}
    \ .
\end{align*}

By applying Bayes' rule and the law of total expectation, note that:
\begin{align*}
    & 
    \EXP\LSS \bigg[ 
     \frac{ \rho^*(1,M,X) }{ \omega^*(1,X) } 
    \mu^* (0,M,X)
    \,\bigg| \, X
    \bigg] 
    \\
    & =
    \int \frac{ \Pr(A=1 \cond M,X) }{ \Pr(A=1 \cond X) }
    \EXP(Y \cond A=0,M,X) f^*(M \cond X) \, dM
    \\
    &
    =
    \int 
    \EXP(Y \cond A=0,M,X) f^*(M \cond A=1,M) \, dM
    \\
    &
    =
    \lambda^*(0,1,X) \ .
    \numeq  \label{eq-proof1-2arm}
\end{align*}

Using this identity, we can reorganize the terms involving $\mu^*(0,M,X)$:
\begin{align*}
    &
    \EXP \LSS
    \Bigg[ 
    \frac{ \rho^*(0,M,X) }{ \widehat{\omega}\LSS(1,X) } 
    \frac{ \widehat{\rho}\LSS(1,M,X) }{ \widehat{\rho}\LSS(0,M,X) } 
    \{ \mu^*(0,M,X) - \widehat{\mu}\LSS(0,M,X) \}
    +
    \frac{ \rho^*(1,M,X) }{ \widehat{\omega}\LSS(1,X) } 
    \widehat{\mu}\LSS(0,M,X)
    \Bigg]
    \\
    &
    =
    \EXP \LSS
    \Bigg[ 
    \bigg\{
    \frac{ \rho^*(0,M,X) }{ \widehat{\omega}\LSS(1,X) } 
    \frac{ \widehat{\rho}\LSS(1,M,X) }{ \widehat{\rho}\LSS(0,M,X) } 
    -
    \frac{ \rho^*(1,M,X) }{ \widehat{\omega}\LSS(1,X) } 
    \bigg\}
    \{ \mu^*(0,M,X) - \widehat{\mu}\LSS(0,M,X) \} 
    \Bigg]
    \\
    &
    \qquad +
    \EXP \LSS
    \Bigg[ 
    \frac{ \rho^*(1,M,X) }{ \widehat{\omega}\LSS(1,X) } 
    \mu^*(0,M,X)
    \Bigg]
    \\
    &
    \stackrel{\eqref{eq-proof1-2arm}}{=}
    \EXP \LSS
    \Bigg[ 
    \frac{ \rho^*(1,M,X) }{ \widehat{\omega}\LSS(1,X) } 
    \bigg\{
    \frac{ \widehat{\rho}\LSS(1,M,X) }{ 
    \rho^*(1,M,X)}
    \frac{\rho^*(0,M,X)}{\widehat{\rho}\LSS(0,M,X) } 
    -
    1
    \bigg\} 
    \{ \mu^*(0,M,X) - \widehat{\mu}\LSS(0,M,X) \} 
    \Bigg]
    \\
    &
    \qquad +
    \EXP \LSS
    \Bigg[ 
    \frac{ 
    \omega^*(1,X)
    }{
    \widehat{\omega}\LSS(1,X)
    }
    \lambda^*(0,1,X)
    \Bigg] \ .
    \numeq  \label{eq-proof3-2arm}
\end{align*}
In addition, we have $\EXP[ \rho^*(1,M,X) \cond X ] = 
    \omega^*(1,X)$, the terms involving $\lambda^*$ reduce to:
    \begin{align*}
    &
    \EXP \LSS
    \Bigg[ 
    -
    \frac{ \rho^*(1,M,X) }{ \widehat{\omega}\LSS(1,X) }  \widehat{\lambda}\LSS(0,1,X) 
    +
    \lambda\LSS(0,1,X)
    - \lambda^*(0,1,X) 
    \Bigg] 
    \\
    &
    =
    \EXP \LSS
    \Bigg[ 
    -
    \frac{ \omega^*(1,X) }{ \widehat{\omega}\LSS(1,X) }  \widehat{\lambda}\LSS(0,1,X)  
    +
    \lambda\LSS(0,1,X)
    - \lambda^*(0,1,X) 
    \Bigg] \ .
    \numeq  \label{eq-proof4-2arm}
    \end{align*}
    
Combining \eqref{eq-proof3-2arm} and \eqref{eq-proof4-2arm}, we obtain the cross-product of estimation errors. Taking the absolute value, we can bound the remainder:
\begin{align*}
    & \big\| \eqref{eq-proof01-2arm} \big\|
    \\
    &
    =
    \left\|
    \EXP \LSS
    \left[ 
    \begin{array}{l}
    \frac{ \rho^*(1,M,X) }{ \widehat{\omega}\LSS(1,X) } 
    \big\{
    \frac{ \widehat{\rho}\LSS(1,M,X) }{ 
    \rho^*(1,M,X)}
    \frac{\rho^*(0,M,X)}{\widehat{\rho}\LSS(0,M,X) } 
    -
    1
    \big\} 
    \{ \mu^*(0,M,X) - \widehat{\mu}\LSS(0,M,X) \} 
    \\
    + 
    \big\{  \frac{ \omega^*(1,X) }{ \widehat{\omega}\LSS(1,X) }  -1 \big\}
    \{  \lambda^*(0,1,X)
    -   \lambda\LSS(0,1,X) \}
    \end{array}
    \right]
    \right\|
    \\
    &
    \lesssim 
    \big\| \widehat{\rho}\LSS(1,M,X) - \rho^*(1,M,X) \big\|_{P,2}
    \big\| \widehat{\mu}\LSS(0,M,X) - \mu^*(0,M,X) \big\|_{P,2}
    \\
    &
    \qquad 
    +
    \big\| \widehat{\omega}\LSS(1,X) - \omega^*(1,X) \big\|_{P,2}
    \big\| \widehat{\lambda}\LSS(0,1,X) - \lambda^*(0,1,X) \big\|_{P,2}
    \\
    &
    = o_P(N^{-1/2}) \ .
    \numeq  \label{eq-proof5-2arm}
\end{align*}
The inequality follows from the Cauchy-Schwarz inequality and Assumption \ref{assumption-reg-2arm}-(i), while the final line follows from the rate conditions in Assumption \ref{assumption-reg-2arm}-(ii).  Thus:
\begin{align*}
    \eqref{term-3-2arm} 
    & = 
    \frac{1}{\sqrt{N}}
\sum_{k=1}^{K} \sum_{i \in \mathcal{I}^{(k)}} 
\big[
    \EXP\LSS \{ \Psi(O_i \con \widehat{\eta}\LSS) \}
    -
    \EXP\LSS \{ \Psi(O_i \con \eta^*) \}
    \big]  
    \\
    &
    \lesssim
    \sqrt{N} 
    \big[ \EXP\LSS \{ \Psi(O \con \widehat{\eta}\LSS) \}
    -
    \EXP\LSS \{ \Psi(O \con \eta^*) \}  \big]
    \\
    & \stackrel{\eqref{eq-proof00-2arm},\eqref{eq-proof5-2arm}}{=} \sqrt{N} o_P(N^{-1/2})
    \\
    &
    = o_P(1) \ .
\end{align*}

\newpage 

Next, we establish that \eqref{term-2-2arm} is $o_P(1)$ by showing $\mathbbm{G}_{k} =o_P(1)$ for each $k$, where
\begin{align*}
\mathbbm{G}_{k} = 
\frac{1}{\sqrt{N/2}}
 \sum_{i \in \mathcal{I}^{(k)}} 
\left[ 
\begin{array}{l}
    \Psi(O_i \con  \widehat{\eta}\LSS)
    -
    \EXP\LSS \{ \Psi(O_i \con  \widehat{\eta}\LSS) \} 
    \\
    -
    \Psi(O_i \con  \eta^*)
    +    
    \EXP\LSS \{ \Psi(O_i \con  \eta^*) \}
\end{array}
    \right] \ .
\end{align*}  

By construction, $\EXP\LSS[ \mathbbm{G}_{k} ] = 0$. Therefore, it suffices to show that $\VAR\LSS[ \mathbbm{G}_{k} ] = o_P(1)$. Specifically, we obtain
\begin{align*}
    &
    \VAR\LSS[ \mathbbm{G}_{k} ]
    \\
    & = 
    \VAR\LSS 
    \left[ 
    \Psi(O_i \con  \widehat{\eta}\LSS)
    -
    \EXP\LSS \{ \Psi(O_i \con  \widehat{\eta}\LSS) \} 
    -
    \Psi(O_i \con  \eta^*)
    +    
    \EXP\LSS \{ \Psi(O_i \con  \eta^*) \} 
    \right]
    \\
    & = 
    \EXP \LSS 
    \left[ 
    \left\{
    \Psi(O_i \con  \widehat{\eta}\LSS)
    -
    \Psi(O_i \con  \eta^*) 
\right\}^2
    \right]
    \\
    & = 
    \EXP \LSS
    \left[ 
    \left\{ 
    \begin{array}{l}
    \frac{ A }{ \widehat{\omega}\LSS(1,X) } 
    \{ Y - \widehat{\lambda}\LSS(1,1,X) \}
    + \lambda\LSS(1,1,X)
    \\
    - \frac{ A }{ \omega^*(1,X) } 
    \{ Y - \lambda^*(1,1,X) \}
    - \lambda^*(1,1,X)
    \\
    -
    \frac{ 1-A }{ \widehat{\omega}\LSS(1,X) } 
    \frac{ \widehat{\rho}\LSS(1,M,X) }{ \widehat{\rho}\LSS(0,M,X) } 
    \{ Y - \widehat{\mu}\LSS(0,M,X) \}
    \\
    +
    \frac{ 1-A }{ \omega^*(1,X) } 
    \frac{ \rho^*(1,M,X) }{ \rho^*(0,M,X) } 
    \{ Y - \mu^*(0,M,X) \}
    \\
    -
    \frac{ A }{ \widehat{\omega}\LSS(1,X) } 
    \{ \widehat{\mu}\LSS(0,M,X) - \widehat{\lambda}\LSS(0,1,X) \}
    \\
    +
    \frac{ A }{ \omega^*(1,X) } 
    \{ \mu^* (0,M,X) - \lambda^*(0,1,X) \}
    \\
    - \lambda\LSS(0,1,X)
    + \lambda^*(0,1,X)
    \end{array}
    \right\}^2
    \right]  
    \\
    &
    \lesssim 
    \sum_{a=0}^{1}
    \big[ 
    \big\| \widehat{\rho}\LSS(a,1,X) - \rho^*(a,1,X) \big\|_{P,2}^2
    +
    \big\| \widehat{\lambda}\LSS(a,a_M,X) - \lambda^*(a,a_M,X) \big\|_{P,2}^2
    \big] 
    \\
    &
    \qquad + \big\| \widehat{\omega}\LSS(1,X) - \omega^*(1,X) \big\|_{P,2}^2
    + \big\| \widehat{\mu}\LSS(0,M,X) - \mu^*(0,M,X) \big\|_{P,2}^2
    \\
    &
    =
    o_P(1) 
    \ .
    \numeq \label{eq-proof3-2arm}
\end{align*}
The inequality follows from the Cauchy-Schwarz inequality and Assumption \ref{assumption-reg-2arm}-(i), while the final line follows from the rate conditions in Assumption \ref{assumption-reg-2arm}-(ii). 

Finally, we establish the consistency of the variance estimator. Note that $\widehat{\sigma}_{\text{SDE}}^2(a_M) = \sum_{k=1}^{K} \widehat{\sigma}_{k,\text{SDE}}^2(a_M)/K$ where
\begin{align*}
    &
    \widehat{\sigma}_{k,\text{SDE}}^2(a_M)
    =
    \frac{1}{ N/2 } 
    \sum_{i \in \mathcal{I}^{(k)} }
    \big\{ 
        \Psi(O_i \con a_M, \widehat{\eta}\LSS)
    -
    \widehat{\text{SDE}}(a_M)
    \big\}^2  \ .
    \end{align*}
It suffices to show the consistency of $\widehat{\sigma}_{k,\text{SDE}}^2(a_M)$, i.e., $\widehat{\sigma}_{k,\text{SDE}}^2(a_M) - \sigma_{\text{SDE}}^2(a_M) = o_P(1)$. We can express this difference as:
\begin{align*}
    &
    \widehat{\sigma}_{k,\text{SDE}}^2(a_M) - \sigma_{\text{SDE}}^2(a_M)
    \\
    &
    =
    \frac{1}{N/2}
    \sum_{i \in \mathcal{I}^{(k)}} 
    \big\{ 
        \Psi(O_i \con a_M, \widehat{\eta}\LSS)
    -
    \widehat{\text{SDE}}(a_M)
    \big\}^2 - \sigma_{\text{SDE}}^2(a_M)
    \\
    &
    =
     \frac{1}{N/2}
    \sum_{i \in \mathcal{I}^{(k)}} 
    \Big[ 
    \big\{ 
        \Psi(O_i \con a_M, \widehat{\eta}\LSS)
    -
    \widehat{\text{SDE}}(a_M)
    \big\}^2
    -
    \big\{ 
        \Psi(O_i \con a_M, \eta^*)
    -
    \text{SDE}^*(a_M)
    \big\}^2
    \Big] 
    \\
    & \qquad 
    +
    \frac{1}{N/2}
    \sum_{i \in \mathcal{I}^{(k)}} 
    \big\{ 
        \Psi(O_i \con a_M, \eta^*)
    -
    \text{SDE}^*(a_M)
    \big\}^2  - \sigma_{\text{SDE}}^2(a_M)
    \\
    &
    = 
    \frac{1}{N/2}
    \sum_{i \in \mathcal{I}^{(k)}} 
    \Big[ 
    \big\{ 
        \Psi(O_i \con a_M, \widehat{\eta}\LSS)
    -
    \widehat{\text{SDE}}(a_M)
    \big\}^2
    -
    \big\{ 
        \Psi(O_i \con a_M, \eta^*)
    -
    \text{SDE}^*(a_M)
    \big\}^2
    \Big]
    + o_P(1) \ .
    \numeq
    \label{eq-proof2}
\end{align*}
The last line holds from the law of large numbers. 

Let $\widehat{\Delta}_i\LSS = \{ 
        \Psi(O_i \con a_M, \widehat{\eta}\LSS)
    -
    \widehat{\text{SDE}}(a_M)
    \}
    -
    \{ 
        \Psi(O_i \con a_M, \eta^*)
    -
    \text{SDE}^*(a_M)
    \}$. Then, \eqref{eq-proof2} can be represented as 
\begin{align*}
    &
    \frac{1}{N/2}
    \sum_{i \in \mathcal{I}^{(k)}} 
    \Big[ 
    \big\{ 
        \Psi(O_i \con a_M, \widehat{\eta}\LSS)
    -
    \widehat{\text{SDE}}(a_M)
    \big\}^2
    -
    \big\{ 
        \Psi(O_i \con a_M, \eta^*)
    -
    \text{SDE}^*(a_M)
    \big\}^2
    \Big]
    \\
    &
    =
    \frac{1}{N/2}
    \sum_{i \in \mathcal{I}^{(k)}} 
\widehat{\Delta}_i\LSS
    \times 
    \big[
\widehat{\Delta}_i\LSS
    +
    2
    \{ 
        \Psi(O_i \con a_M, \eta^*)
    -
    \text{SDE}^*(a_M)
    \}
    \big]
    \\
    &
    =
    \frac{1}{N/2}
    \sum_{i \in \mathcal{I}^{(k)}} 
\{ \widehat{\Delta}_i\LSS \}^2
+
    \frac{2}{N/2}
    \sum_{i \in \mathcal{I}^{(k)}} 
    \widehat{\Delta}_i\LSS 
    \{ 
        \Psi(O_i \con a_M, \eta^*)
    -
    \text{SDE}^*(a_M)
    \}  
    \ .
\end{align*}
 From the H\"older's inequality, we find the absolute value of \eqref{eq-proof2} is bounded by
 \begin{align*}
     &
     \big\| \eqref{eq-proof2} \big\|
     \\
     &
     \leq 
         \frac{1}{N/2}
    \sum_{i \in \mathcal{I}^{(k)}} 
\{ \widehat{\Delta}_i\LSS \}^2
+
2
\sqrt{
    \frac{1}{N/2}
    \sum_{i \in \mathcal{I}^{(k)}} 
    \{ \widehat{\Delta}_i\LSS \}^2
    }
    \underbrace{
    \sqrt{
    \frac{1}{N/2}
    \sum_{i \in \mathcal{I}^{(k)}} 
    \{ 
        \Psi(O_i \con a_M, \eta^*)
    -
    \text{SDE}^*(a_M)
    \}  ^2
    } }_{\sigma^2+o_P(1)=O_P(1)}
    \ .
 \end{align*} 
 Therefore, \eqref{eq-proof2} is $o_P(1)$ if $\frac{1}{N/2}
    \sum_{i \in \mathcal{I}^{(k)}} 
    \{ \widehat{\Delta}_i\LSS \}^2 = o_P(1)$. From some algebra, we find
    \begin{align*}
    &
    \frac{1}{N/2}
    \sum_{i \in \mathcal{I}^{(k)}} 
    \{ \widehat{\Delta}_i\LSS \}^2
    \\
    &
    =
    \frac{1}{N/2}
    \sum_{i \in \mathcal{I}^{(k)}} 
    \Big[ 
    \{ 
    \Psi(O_i \con a_M, \widehat{\eta}\LSS)
    -
    \widehat{\text{SDE}}(a_M)
    \}
    -
    \{ 
        \Psi(O_i \con a_M, \eta^*)
    -
    \text{SDE}^*(a_M)
    \}
    \Big]^2
    \\
    &
    \lesssim 
    \frac{1}{N/2}
    \sum_{i \in \mathcal{I}^{(k)}}  
    \{ 
    \Psi(O_i \con a_M, \widehat{\eta}\LSS)
    -
        \Psi(O_i \con a_M, \eta^*)
    \}^2
    +
    \{ \widehat{\text{SDE}}(a_M) - \text{SDE}^*(a_M) \}^2
    \\
    &
    =
    \EXP\LSS
    \big[ 
    \{ 
    \Psi(O_i \con a_M, \widehat{\eta}\LSS)
    -
        \Psi(O_i \con a_M, \eta^*)
    \}^2
    \big] 
    +
    o_P(1)
    \\
    &
    =
    o_P(1) \ .
    \end{align*}
The fourth line holds from $\widehat{\text{SDE}}(a_M) - \text{SDE}^*(a_M) = o_P(1) $ and the law of large numbers. The last line holds from \eqref{eq-proof3-2arm}. This concludes the proof.

\vspace{\fill}\clearpage

\section{Details of Section \protect\ref{sec: falsification SEE}} \label{app:third estimator}

The separable effects estimated by $\widetilde{\text{SDE}}_{[2]}$ and $\widetilde{\text{SIE}}_{[2]}$ are represented as
\begin{align*}
    \EXP[Y^{a_Y=1,a_M}-Y^{a_Y=0,a_M} \cond A_Y=A_M]
    =
    \theta^{*}(1,a_M)
    -
    \theta^{*}(0,a_M)
    \ , 
    \\
    \EXP[Y^{a_Y,a_M=1}-Y^{a_Y,a_M=0} \cond A_Y=A_M] 
    =
    \theta^{*}(a_Y,1)
    -
    \theta^{*}(a_Y,0)
    \ ,
\end{align*}
where $\theta^{*}(a_Y,a_M) \equiv \EXP[ Y^{a_Y,a_M} \cond A_Y=A_M]$. These effects are identical to $\text{SDE}_{[2]}(a_M)$ and $\text{SIE}_{[2]}(a_Y)$ if \eqref{eq-falsification mean 2} and \eqref{eq-falsification mean 3} hold.

We suppress the subscript $_{[4]}$ for notational brevity. Therefore, given an estimator $\widetilde{\theta}(a_Y,a_M)$ for $\theta^*(a_Y,a_M)$, the corresponding estimators for the separable effects can be constructed as  $\widetilde{\text{SDE}}_{[2]}(a_M) = \widetilde{\theta}(1,a_M)
    -
    \widetilde{\theta}(0,a_M)$ and $\widetilde{\text{SIE}}_{[2]}(a_Y)= 
    \widetilde{\theta}(a_Y,1) - \widetilde{\theta}(a_Y,0)$. Therefore, it suffices to focus on $\theta^*(a_Y,a_M)$.

We first derive the EIF for $\theta^*(a_Y,a_M)$. Let $\mathcal{M}$ denote the nonparametric model of the four-arm data $O=(Y,A_Y,A_M,M,X)$, and let $\mathcal{M}_{\text{sub}} = \{ P_t \cond t \in \R \}$ be the regular parametric submodel of $\mathcal{M}$ indexed by a one-dimensional parameter $t \in \R$. Without loss of generality, we assume that the true data-generating law $P^* \in \mathcal{M}$ is recovered at $P_{t=0} \in \mathcal{M}_{\text{sub}}$. Let $\EXP_{t}$ be the expectation under $P_t$, $f_t(O)$ be the density of $P_t$ and $s_t(O) = \nabla_{t} f_t(O)/f_t(O)$ be the score function. Furthermore, we denote  
$\nu_t(a_Y,a_M,X) = \EXP_t(Y \cond A_Y=a_Y,A_M=a_M,X)$ and
$\pi_t(a_Y,a_M,X)=\Pr_{t}(A_Y=a_Y,A_M=a_M \cond X)$. The asterisk notation $^*$ indicates that a function is evaluated at $P^*=P_0$, e.g., $\nu^*$ and $\pi^*$.

Under $P^*$, we have $\theta^{*}(a_Y,a_M)=\theta_N^{*}(a_Y,a_M)/\theta_M^{*}(a_Y,a_M)$ where
\begin{align*}
   &
   \theta_N^{*}(a_Y,a_M) 
   =
   \EXP [
        \nu^*(a_Y,a_M,X)
        \ind (A_Y=A_M)
    ] \ ,
    \\
    &
    \theta_M^{*}(a_Y,a_M)
    =
    \Pr(A_Y=A_M) \ .
\end{align*}
Similarly, under $P_t$, we have
\begin{align*}
    &
    \theta_{t}(a_Y,a_M)
    =
    \frac{\theta_{t,N}(a_Y,a_M) }{\theta_{t,D}(a_Y,a_M) }
    =
    \frac{
    \EXP_{t} [
    \nu_t(a_Y,a_M,X)
        \ind (A_Y=A_M)
    ] }{ 
    \Pr_{t}(A_Y=A_M)
    } \ .
\end{align*}

We first show the an influence function for the numerator $\theta_{t,N}(a_Y,a_M)$ in $\mathcal{M}$. The pathwise derivative of the numerator is
\begin{align*}
    &
    \nabla_{t}
    \theta_{t,N}(a_Y,a_M)
    \\
    &
    =
    \EXP_t 
    \left[ 
    \begin{array}{l}
        s_t(O)
        \nu_t(a_Y,a_M,X)
        \ind (A_Y=A_M)
        \\ 
        +
        \EXP_{t}\{ Y s_t(Y \cond A_Y,A_M,X) \cond A_Y=a_Y,A_M=a_M,X\}
        \ind (A_Y=A_M)
    \end{array}
    \right] 
    \\
    &
    =
    \EXP_t 
    \left[ 
    \begin{array}{l}
        s_t(O)
        \nu_t(a_Y,a_M,X)
        \ind (A_Y=A_M)
        \\ 
        +
        \EXP_{t}[ \{Y-\nu_t(a_Y,a_M,X) \} s_t(Y \cond A_Y,A_M,X) \cond A_Y=a_Y,A_M=a_M,X]
        \ind (A_Y=A_M)
    \end{array}
    \right] 
    \\
    &
    =
    \EXP_t 
    \left[ 
    \begin{array}{l}
        s_t(O)
        \nu_t(a_Y,a_M,X)
        \ind (A_Y=A_M)
        \\ 
        +
        \EXP_{t}[ \{Y-\nu_t(a_Y,a_M,X) \} s_t(O) \cond A_Y=a_Y,A_M=a_M,X]
        \ind (A_Y=A_M)
    \end{array}
    \right]  
    \\
    &
    =
    \EXP_t 
    \left[ 
    \begin{array}{l}
        s_t(O)
        \nu_t(a_Y,a_M,X)
        \ind (A_Y=A_M)
        \\ 
        +
        \EXP_{t}[ \{Y-\nu_t(a_Y,a_M,X) \}
        \Pr_{t}(A_Y=A_M \cond X)
        s_t(O) \cond A_Y=a_Y,A_M=a_M,X] 
    \end{array}
    \right] 
    \\
    &
    =
    \EXP_t 
    \Bigg[
        s_t(O)
        \underbrace{
        \Bigg[
    \begin{array}{l}
    \displaystyle{
    \frac{ \ind(A_Y=a_Y,A_M=a_M) \{Y-\nu_t(a_Y,a_M,X) \}
        \Pr_{t}(A_Y=A_M \cond X) }{
        \pi_t(a_Y,a_M,X) }
        }
        \\
        +
        \nu_t(a_Y,a_M,X)
        \ind (A_Y=A_M) 
    \end{array}
    \Bigg] }_{\equiv \mathcal{D}_{t,N}(O; a_Y,a_M)}
    \Bigg] \ .
\end{align*}

Next, we derive $\nabla_{t}\theta_{t,D}(a_Y,a_M)$:
\begin{align*}
    \nabla_t \theta_{t,D}(a_Y,a_M) = \EXP_t[s_t(O) \ind(A_Y=A_M)] \ .
\end{align*}

Now we combine these pieces. Using the quotient rule for pathwise derivatives:
\begin{align*}
    & \nabla_t \theta_t(a_Y,a_M) 
    \\
    &
    = \frac{\theta_{t,D}(a_Y,a_M) \nabla_t \theta_{t,N}(a_Y,a_M) - \theta_{t,N}(a_Y,a_M) \nabla_t \theta_{t,M}(a_Y,a_M)}{\theta_{t,M}(a_Y,a_M)^2} 
    \\
    &
    = \frac{\nabla_t \theta_{t,N}(a_Y,a_M) - \theta_t(a_Y,a_M) \nabla_t \theta_{t,M}(a_Y,a_M) }{\theta_{t,M}(a_Y,a_M)} 
    \ .
\end{align*}

We can rewrite this as an expectation with the score $s_t(O)$:
\begin{align*}
    \nabla_t \theta_t(a_Y,a_M) = \EXP_t \Bigg[ s_t(O) \underbrace{ \frac{\mathcal{D}_{t,N}(O; a_Y,a_M) - \theta_t(a_Y,a_M) \ind(A_Y=A_M)}{\theta_{t,M}(a_Y,a_M)} }_{=\varphi_t(O; a_Y,a_M)} \Bigg] \ ,
\end{align*}
where $\varphi_t(O; a_Y,a_M)$ is given by
\begin{align*}
    \varphi_t(O; a_Y,a_M)
    =
    \frac{
    \left[
    \begin{array}{l}
    \displaystyle{
    \frac{ \ind(A_Y=a_Y,A_M=a_M) \{Y-\nu_t(a_Y,a_M,X) \}
        \Pr_{t}(A_Y=A_M \cond X) }{
        \pi_t(a_Y,a_M,X) }
        }
        \\
        +
        \nu_t(a_Y,a_M,X)
        \ind (A_Y=A_M) 
    -
    \theta_t(a_Y,a_M) \ind (A_Y=A_M)
    \end{array}
    \right]}{
    \Pr_t(A_Y=A_M)
    } \ .
\end{align*}
It is straightforward to demonstrate that $\EXP_{t} [ \varphi_t(O; a_Y,a_M) ] =0$. Therefore, at $t=0$, we have
\begin{align*}
    \nabla_t \theta_t \cond_{t=0}
    =
    \EXP [ s^*(O) \varphi^*(O; a_Y,a_M) ]
\end{align*}
implying that the following $\varphi^*(O; a_Y,a_M)$ is an influence function for $\theta^*(a_Y,a_M)$ in the nonparametric model $\mathcal{M}$.
\begin{align*}
    \varphi^*(O; a_Y,a_M)
    =
    \frac{
    \varphi_1(O \con a_Y,a_M, \gamma^*)
    -
    \theta^*(a_Y,a_M) \ind (A_Y=A_M)}{
    \Pr(A_Y=A_M)
    } \ , 
\end{align*}
where
\begin{align*}
    & \varphi_1(O \con a_Y,a_M, \gamma^*)
    \\
    &
    =
    \displaystyle{
    \frac{ \ind(A_Y=a_Y,A_M=a_M) \{Y-\nu^*(a_Y,a_M,X) \}
        \Pr(A_Y=A_M \cond X) }{
        \pi^*(a_Y,a_M,X) }
        } 
        +
        \nu^*(a_Y,a_M,X)
        \ind (A_Y=A_M) 
\end{align*}
with $\gamma^*=(\nu^*,\Pr(A_Y=A_M | X), \pi^*)$. Since the model is nonparametric, the influence function $\varphi^*(O; a_Y,a_M)$ is unique, therefore it is the EIF for $\theta^*(a_Y,a_M)$.

The EIF-based cross-fitted estimator for $\theta^*(a_Y,a_M)$ is given by:
\begin{align*}
    \widetilde{\theta}(a_Y,a_M)
    =
    \frac{  
    \sum_{k=1}^{K}
    \sum_{i\in \mathcal{I}^{(k)}}
    \varphi_1(O_i \con a_Y,a_M, \widehat{\gamma}\LSS) }{ 
    \sum_{k=1}^{K}
    \sum_{i\in \mathcal{I}^{(k)}}
    \ind(A_{Y,i}=A_{M,i})
    } \ ,
    \numeq \label{eq-theta estimator}
\end{align*}
where $\widehat{\gamma}\LSS$ is a nonparametric estimator for $\gamma^*$ trained on $\mathcal{I}^{(-k)}$.  

To test the falsification conditions in \eqref{eq-falsification mean 2} and \eqref{eq-falsification mean 3}, we can construct an estimator based on the EIF. For instance, the condition in \eqref{eq-falsification mean 2} is equivalent to the null hypothesis $H_0: \Xi^* = 0$, where we define the parameter $\Xi^*$ as
\begin{align} \label{eq-falsification mean 2-alternative} 
    \Xi^*
    =
    \theta^{*}(1,a_M)
    -
    \theta^{*}(0,a_M)
    -
    \text{SDE}_{[2]}(a_M) \ .
\end{align}
The EIF for $\Xi^*$ is given by the difference between the EIFs of its components:
\begin{align*}
    &
    \varphi^*(O; 1,a_M)
    -
    \varphi^*(O; 0,a_M)
    -
    \frac{ 
    \{
    \psi(O \con 1, a_M, \eta^*)
    -
    \psi(O \con 0, a_M, \eta^*)
    -
    \text{SDE}_{[2]}(a_M)
    \} \ind(A_Y=A_M)
    }{ \Pr(A_Y=A_M) }
    \\
    &
    =
    \frac{
    \varphi_1(O \con 1,a_M, \gamma^*)
    -
    \varphi_1(O \con 0,a_M, \gamma^*)
    -
    \{ \theta^*(1,a_M) - \theta^*(0,a_M) \} \ind (A_Y=A_M)}{
    \Pr(A_Y=A_M)
    }    
    \\
    &
    \qquad 
    -
    \frac{ 
    \{
    \psi(O \con 1, a_M, \eta^*)
    -
    \psi(O \con 0, a_M, \eta^*)
    -
    \text{SDE}_{[2]}(a_M)
    \} \ind(A_Y=A_M)
    }{ \Pr(A_Y=A_M) }
    \ .
\end{align*}
where $\psi$ is defined in \eqref{eq-psi}. The factor $\ind(A_Y=A_M)/\Pr(A_Y=A_M)$ in
$\psi(O \con 1, a_M, \eta^*) - \psi(O \con 0, a_M, \eta^*) - \text{SDE}_{[2]}(a_M)$
is included because the two-arm data represent a strict subset of the four-arm data where $A = A_Y = A_M$.

We can estimate $\Xi^*$ by plugging in the respective estimators:
\begin{align*}
    \widehat{\Xi} = \widetilde{\theta}(1,a_M) - \widetilde{\theta}(0,a_M) - \widehat{\text{SDE}}_{[2]}(a_M) \ ,
\end{align*}
where $\widetilde{\theta}(a_Y,a_M)$ and $\widehat{\text{SDE}}_{[2]}(a_M)$ are defined in \eqref{eq-theta estimator} and \eqref{eq-2arm estimator}, respectively. 

The corresponding consistent variance estimator, $\widehat{\sigma}_{\Xi}^2$, is the empirical variance of the estimated EIF, i.e., 
\begin{align*}
    &
    \widehat{\sigma}_{\Xi}^2
    \\
    &
    =
    \frac{1}{N} \sum_{k=1}^{K} \sum_{i \in \mathcal{I}^{(k)}}  
    \left[
    \frac{ 
    \begin{array}{l}
        \varphi_1(O \con 1,a_M, \widehat{\gamma}^{(-k)})
    -
    \varphi_1(O \con 0,a_M, \widehat{\gamma}^{(-k)})
    -
    \{ \widetilde{\theta}(1,a_M) - \widetilde{\theta}(0,a_M) \} \ind (A_{Y,i}=A_{M,i})
    \\
    -
    \{
    \psi(O \con 1, a_M, \widehat{\eta}\LSS)
    -
    \psi(O \con 0, a_M, \widehat{\eta}\LSS)
    -
    \widehat{\text{SDE}}_{[2]}(a_M) 
    \}  \ind (A_{Y,i}=A_{M,i})
    \end{array} }{
    \displaystyle{ 
    \frac{1}{N}
    \sum_{k=1}^{K}
    \sum_{i\in \mathcal{I}^{(k)}}
    \ind(A_{Y,i}=A_{M,i}) } } 
    \right] ^2  \ .
\end{align*}
Assuming the nuisance functions are estimated at a rate of $o_P(N^{-1/4})$, we have:
\begin{align*}
    \sqrt{N} \,
    \frac{ \widehat{\Xi} - \Xi^* }{ \widehat{\sigma}_{\Xi}  }
    \stackrel{D}{\rightarrow}  N(0,1) \ .
\end{align*}
Consequently, we can use the Wald statistic, $W = \sqrt{N} | \widehat{\Xi} | / \widehat{\sigma}_{\Xi}$, to test the null hypothesis $H_0: \Xi^* = 0$.

\vspace{\fill}\clearpage

\section{Simulation Study: Details of $\widehat{\text{SDE}}_{[2]}$ and $\widehat{\text{SIE}}_{[2]}$} \label{app:second estimator}

The S-learner (single-learner) strategy includes treatment as a predictor and uses the entire training dataset, denoted by $\mathcal{I}^{(-k)}$. Specifically, to estimate the outcome regression $\mu^*(a,M,X) = \EXP(Y \cond A=a, M, X)$, one can fit a machine learning model predicting $Y$ from $(A, M, X)$. We denote this S-learner model with the superscript ($S$): 
\begin{align*}
    \widehat{\mu}^{(S)} (\cdot) = \texttt{ML}(Y_i \sim A_i + M_i + X_i, i \in \mathcal{I}^{(-k)})
\end{align*}
The estimated outcome regression for $A=a$ is obtained by predicting $\widehat{\mu}^{(S)}$ with $A$ fixed at $a$, i.e., $\widehat{\mu}^{(S)}(A=a, M, X)$.

Next, $\lambda^*(a,a',X) = \EXP\{\mu^*(a,M,X) \cond A=a', X\}$ can be estimated by fitting the following model: 
\begin{align*}
    \widehat{\lambda}^{(S)} (a,\cdot) = \texttt{ML}(\widehat{\mu}_i^{(S)}(a) \sim A_i + X_i, i \in \mathcal{I}^{(-k)})
\end{align*}
where $\widehat{\mu}_i^{(S)}(a) = \widehat{\mu}^{(S)}(a, M_i, X_i)$ for $i \in \mathcal{I}^{(-k)}$. The final estimate of $\lambda^*(a,a',X)$ is obtained by predicting from $\widehat{\lambda}^{(S)}$ at $(A=a', X)$.

In contrast, the T-learner (two-learner) strategy does not include treatment $A$ as a predictor. Instead, it splits the training data into two sets: $\mathcal{I}^{(-k)}_1$ with $A=1$ and $\mathcal{I}^{(-k)}_0$ with $A=0$. We denote this T-learner model with the superscript ($T$), and the corresponding outcome regression is defined as:
\begin{align*}
    \widehat{\mu}^{(T)} (a,\cdot) = \texttt{ML}(Y_i \sim M_i + X_i, \in  \mathcal{I}^{(-k)}_{a}) \ , \quad a \in \{0,1\} \ ,
\end{align*}
which estimates $\mu^*(a,M,X)$ using only the units under $A=a$. Here, $A$ is not used explicitly as a predictor, as it determines the training data partitions. The estimated outcome regression for $A=a$ is obtained by predicting from $\widehat{\mu}^{(T)}(a,\cdot)$ at $(M, X)$.  A T-learner for $\lambda^*$ is then obtained by fitting a machine learning model predicting $\widehat{\mu}(a)$ from $X$ using the subset corresponding to $A=a'$:
\begin{align*}
    \widehat{\lambda}^{(T)} (a,a',\cdot) = \texttt{ML}(\widehat{\mu}_i^{(T)}(a) \sim X_i, i \in \mathcal{I}^{(-k)}_{a'})
    \ , \quad (a,a') \in \{0,1\}^{\otimes 2} \ ,
\end{align*}
where $\widehat{\mu}_i^{(T)}(a) = \widehat{\mu}^{(T)}(a, M_i, X_i)$ for $i \in \mathcal{I}^{(-k)}_{a'}$.
The final estimate of $\lambda^*(a,a',X)$ is then obtained by predicting from $\widehat{\lambda}^{(T)}(a,a',\cdot)$ at $X$.

Following the estimation strategy, we can aggregate the two estimated influence functions from the S- and T-learners to obtain an ensemble estimator for the SDE and SIE. For example, the ensemble SDE estimator can be obtained by 
\begin{align*}
    \widehat{\text{SDE}}(a_M)
    =
    \frac{1}{N}
    \sum_{k=1}^{K}
    \sum_{i \in \mathcal{I}^{(k)} }
    \frac{1}{2}
    \left\{
    \begin{array}{l}
        \psi(O_i \con 1,a_M, \widehat{\eta}^{(-k),S})
    -
    \psi(O_i \con 0,a_M,  \widehat{\eta}^{(-k),S})
    \\
       + \psi(O_i \con 1,a_M, \widehat{\eta}^{(-k),T})
    -
    \psi(O_i \con 0,a_M,  \widehat{\eta}^{(-k),T})
    \end{array}
    \right\} \ , 
\end{align*}
where $\widehat{\eta}^{(-k),\star}$ denotes the estimated nuisance functions obtained from either the S-learner or T-learner approach.

\vspace{\fill}\clearpage

\section{Description of Variables}\label{app:descri}

\begin{table}[h]
\centering
\begin{threeparttable}
\caption{Description of Variables}
\label{tab:my_label}    
\small{\begin{tabular}{L{3.5cm}L{3cm}L{8.5cm}}
\hline    
Type     & Variables & Description \\
\hline    
Outcome $Y$ & Test scores &  A student's test scores in a test block ranging from 0 to 25.\\
Treatment Component $A_M$ & Extra time & Whether a student received extra time (No = 0; Yes = 1)\\
Treatment Component  $A_Y$ & Separate session & Whether a student received a separate session (No = 0; Yes = 1)\\
Mediator $M_k$ & Item access & A student's cumulative number of accessed test items by a certain time point $k$ \\   
{\footnotesize Pre-treatment} & Sex & A student's sex (Male = 0; Female =1)\\ 
Covariate $X$ & Race & A student's race/ethnicity (White = 0; African American = 1; Hispanic = 2; Others = 3)\\
& PARED & Parents' education level (High school =0; Grad college = 1; I do not know = 2; Omitted and missing = 3)\\
& IEP & Whether a student has disabilities or not (No = 0; Yes = 1)\\
& NSLP eligibility & Whether a student is eligible for National School Lunch Program (Others = 0; Eligible = 1)\\
& Planned math course & A student's math course planned to take next year (Basic Math = 0; Algebra I = 1; Geometry = 2; Algebra II = 3; I do not know = 4; Omitted and missing = 5)\\
\hline         
\end{tabular}
}
\vspace{0.05in}
\begin{tablenotes}[para,flushleft]
\footnotesize %NOTE: ETA = Extended time accommodation. \\
SOURCE: U.S. Department of Education, National Center for Education Statistics (NCES), 2017 NAEP Grade 8 Mathematics Process Data, Student Features Data File Partial Form and Response Data File.
\end{tablenotes}
\end{threeparttable}    
\end{table}

\newpage

\end{document}